\documentclass[lettersize,journal]{IEEEtran}
\usepackage{amsmath,amsfonts}
\usepackage{algorithmic}
\usepackage{algorithm}
\usepackage{array}
\usepackage{booktabs}
\usepackage{bm}  
\usepackage{fancyhdr}
\usepackage{makecell}
\usepackage{multirow}
\usepackage[caption=false,font=normalsize,labelfont=sf,textfont=sf]{subfig}
\usepackage{textcomp}
\usepackage{stfloats}
\usepackage{url}
\usepackage{verbatim}
\usepackage{graphicx}
\usepackage{cite}
\begin{document}

\title{Force-IMU Fusion-Based Sensing Acupuncture Needle and Quantitative Analysis System for Acupuncture Manipulations}

\author{Peng Tian, Kang Yu, Tianyun Jiang, Yuqi Wang,~\IEEEmembership{Student Member,~IEEE}, Haiying Zhang, Hao Yang, \\Yunfeng Wang, Jun Zhang, Shuo Gao,~\IEEEmembership{Senior Member,~IEEE}, and Junhong Gao  % <-this % stops a space
\thanks{This work was supported by the National Key R\&D Program of China (No. 2024YFC3505200, 2024YFC3505205) and the Fundamental Research Funds for the Central Public Welfare Research Institutes (No. ZZ-JQ2023007). \textit{(Corresponding author: Yunfeng Wang, Jun Zhang, Shuo Gao, Junhong Gao.)}}% <-this % stops a space
\thanks{Peng Tian, Yuqi Wang, Haiying Zhang, Hao Yang, Yunfeng Wang, and Jun Zhang are with the Institute of Microelectronics of Chinese Academy of Sciences, Beijing 100029, China and the University of Chinese Academy of Sciences, Beijing 100049, China (e-mail:  tianpeng@ime.ac.cn; wangyuqi24@ime.ac.cn; zhanghaiying@ime.ac.cn; yanghao@ime.ac.cn; wangyunfeng@ime.ac.cn; zhangjun@ime.ac.cn).}% <-this % stops a space
\thanks{Kang Yu is with the Institute of Microelectronics of Chinese Academy of Sciences, Beijing 100029, China (e-mail: yukang@ime.ac.cn)}
\thanks{Tianyun Jiang, and Junhong Gao are with Institute of Acupuncture and Moxibustion, China Academy of Chinese Medical Sciences, Beijing 100700, China (e-mail: tianyunjiang@hotmail.com; gaojh@mail.cintcm.ac.cn)}
\thanks{Shuo Gao is with the School of Instrumentation Science and Optoelectronic Engineering, Beihang University, Beijing 100191, China (e-mail: shuo\_gao@buaa.edu.cn)}}

\markboth{Journal of \LaTeX\ Class Files,~Vol.~14, No.~8, August~2021}%
{Shell \MakeLowercase{\textit{et al.}}: A Sample Article Using IEEEtran.cls for IEEE Journals}


\maketitle

\begin{abstract}
Acupuncture, one of the key therapeutic methods in Traditional Chinese Medicine (TCM), has been widely adopted in various clinical fields. Quantitative research on acupuncture manipulation parameters is critical to achieve standardized techniques. However, quantitative mechanical detection of acupuncture parameters remains limited. This study establishes a kinematic and dynamic model of acupuncture, identifying key parameters such as lifting-thrusting force, acceleration, velocity, displacement, as well as twirling-rotating angular velocity and angle. To measure these critical parameters, we propose a quantitative system comprising a sensing needle equipped with a force sensor and an inertial measurement unit (IMU), as well as an external camera module to capture image information. By fusing visual and IMU data, we accurately identify the stationary or motion states of the needle, enabling segmented computation of lifting-thrusting velocity and displacement. The experimental results demonstrate that the sensing needle achieves comprehensive detection with high precision, featuring a nonlinearity error of 0.45\% in force measurement and an RMSE of 1.2 mm in displacement. The extracted parameters provide an objective description of the operational characteristics and motion patterns of the four basic acupuncture manipulations. These findings provide valuable tools and methods for research in acupuncture standardization.
\end{abstract}


\section{Introduction}
\thispagestyle{fancy}
Acupuncture, as an integral component of Traditional Chinese Medicine, has gained widespread application in clinical practice due to its significant therapeutic effects \cite{ref1, ref2, ref3}. Acupuncture manipulations are indispensable in both acupuncture education and clinical treatment, playing a pivotal role in achieving therapeutic outcomes [4]. The quantification of acupuncture manipulations requires the identification and characterization of their key parameters. The primary approaches to acupuncture quantification are computational modeling, force measurement, and motion detection techniques. Specifically, the lifting-thrusting velocity and displacement are critical parameters in the computational model [5], and their precise measurement is essential for refining and validating the model. Furthermore, force measurement and motion detection techniques enable the quantification of applied forces and subtle motion changes during acupuncture. These methodologies have demonstrated feasibility in applications such as phantom-based models [6] and human-computer interaction systems for virtual acupuncture [7].

Despite the widespread application of various quantification methods, such as force sensing, visual perception, and posture gloves, several unresolved issues remain. The first issue is the development of a mechanical model to describe translation and rotation during acupuncture procedures, i.e., the kinematic and dynamic models of acupuncture. This model helps to explain the physical relationships between key parameters of acupuncture and to reveal the mechanical characteristics of different acupuncture manipulations. The second is the design of miniaturized sensing acupuncture needle to directly and comprehensively measure needling parameters, including challenging parameters such as lifting-thrusting velocity and displacement, which are difficult to obtain during acupuncture.

Acupuncture mechanical models play a significant role in elucidating the mechanism of acupuncture manipulation and analyzing key parameters. Current research indicates that researchers have modeled the mechanical characteristics of acupuncture needles and biological tissue \cite{ref19, ref20, ref21} and analyzed interaction relationships between different surfaces of needles and acupuncture points [8]. It is noting that acupuncture mechanics encompasses not only three degrees of freedom in the kinematics of lifting-thrusting operations but also three degrees of freedom in the dynamics of twirling-rotating operations. Current model investigations encounter challenges related to insufficient degrees of freedom, and the simplification of dynamic models may compromise system robustness.

In the detection of acupuncture force and torque, methods typically involve integrating a fiber Bragg grating (FBG) [9] or strain gauge \cite{ref10, ref11} between the needle body and handle to measure the force and torque applied by the physician to the needle [12]. However, relying solely on these signals for cycle segmentation introduces inaccuracies in the detection of the acupuncture manipulation cycle. Such inaccuracies arise due to the influence of the reactive force exerted by biological tissues on the needle, as well as other related factors such as force signal filtering, which can affect the precision of period calculation. These challenges highlight the potential advantages of cycle segmentation methods based on displacement measurements, which exhibit distinct advantages in accurately identifying the needling cycle. This is because displacement represents the cumulative effect of both the applied force and the reactive force exerted by the tissue, providing a more comprehensive and reliable basis for cycle segmentation. However, while methods employing optical sensors to acquire needle displacement offer high accuracy in calculation, their practical application is limited by their relatively large size, which poses limitations in clinical practice [13].

\begin{figure*}[t]
\centering
\includegraphics[width=\linewidth]{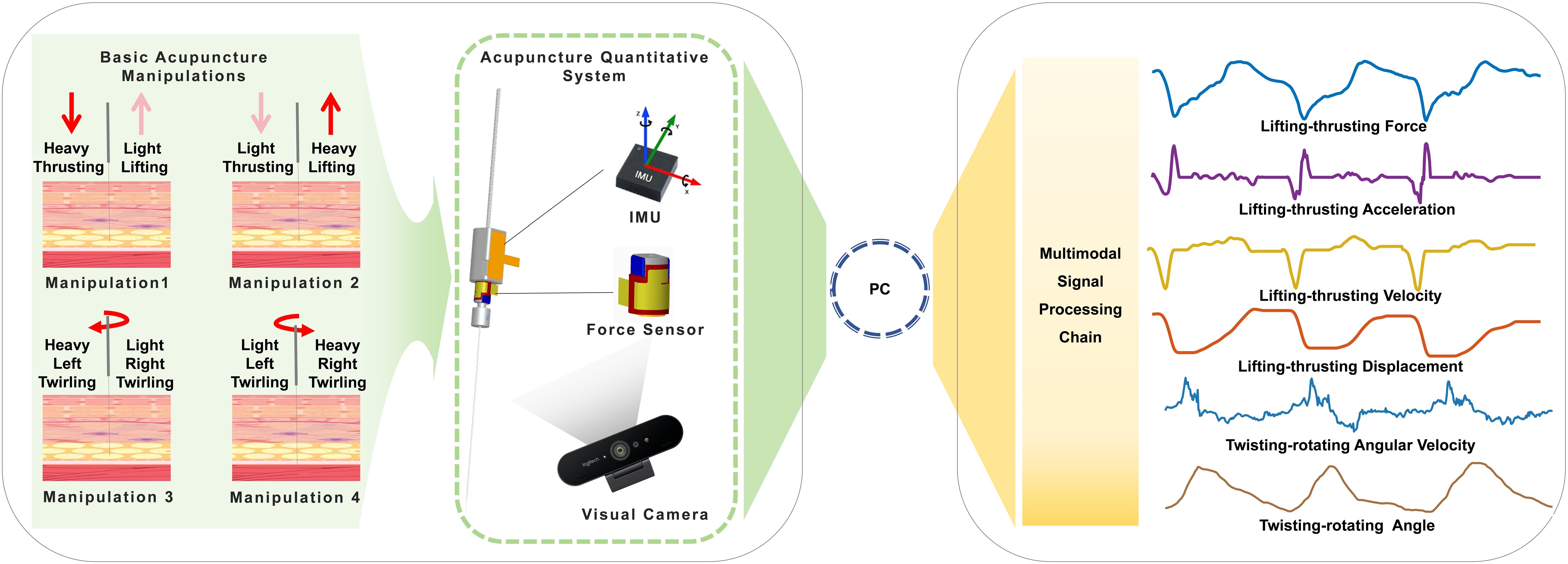}
\caption{The proposed acupuncture quantitative system. The system comprises force, IMU sensors, and a optical camera. The key parameters, including lifting-thrusting force, acceleration, velocity, displacement, as well as twirling-rotating angular velocity and angle, are calculated by a multimodal signal processing chain.}
\label{fig_1}
\end{figure*}

In terms of motion parameter detection, the gesture glove equipped with an inertial measurement unit (IMU) represents an indirect method to monitor needle movement. Even when acupuncture manipulations are consistent, indirect methods may introduce discrepancies in measurement accuracy.
 To address the aforementioned issues, an acupuncture quantitative system is proposed (Fig.1). Based on the analysis of acupuncture dynamics, a six-degree-of-freedom kinematic and dynamic model of acupuncture has been established. Within this model, essential parameters requiring detection during acupuncture procedures have been identified, including lifting-thrusting force, acceleration, velocity, displacement, as well as twirling-rotating angular velocity and angle. Therefore, a sensing acupuncture needle capable of detecting the six parameters (Force, Acceleration, Angular velocity, Angle, Velocity, and Displacement, FAVD) is proposed and named FAVD-Acupuncture Sensor (FAVD-Acusensor). Using the FAVD-Acusensor based on force-IMU fusion, the six critical parameters mentioned above can be detected. To address the errors associated with calculating velocity and displacement from acceleration signals in the IMU, visual recognition is utilized for needle stationary/motion state identification, and a segmented integration method is used to improve the accuracy of velocity and displacement calculations. Compared to state-of-the-art research, the proposed method enables the direct and comprehensive measurement of key parameters, enabling the calculation of lifting-thrusting velocity and displacement during acupuncture, which have been challenging to measure. The method provides a systematic characterization of the kinematic and dynamic parameters of acupuncture manipulations. Building on this foundation, the system can be extended to applications such as standardized training for acupuncture education and trajectory optimization for acupuncture robotic systems, demonstrating significant clinical and technological application potential \cite{ref14, ref15}. Prior to development, we conducted a comprehensive review of existing literature to identify key characteristics, as detailed in the following section.

\section{Overview Of Previous Work}

\begin{figure}[t]
\centering
\includegraphics[width=0.75\linewidth]{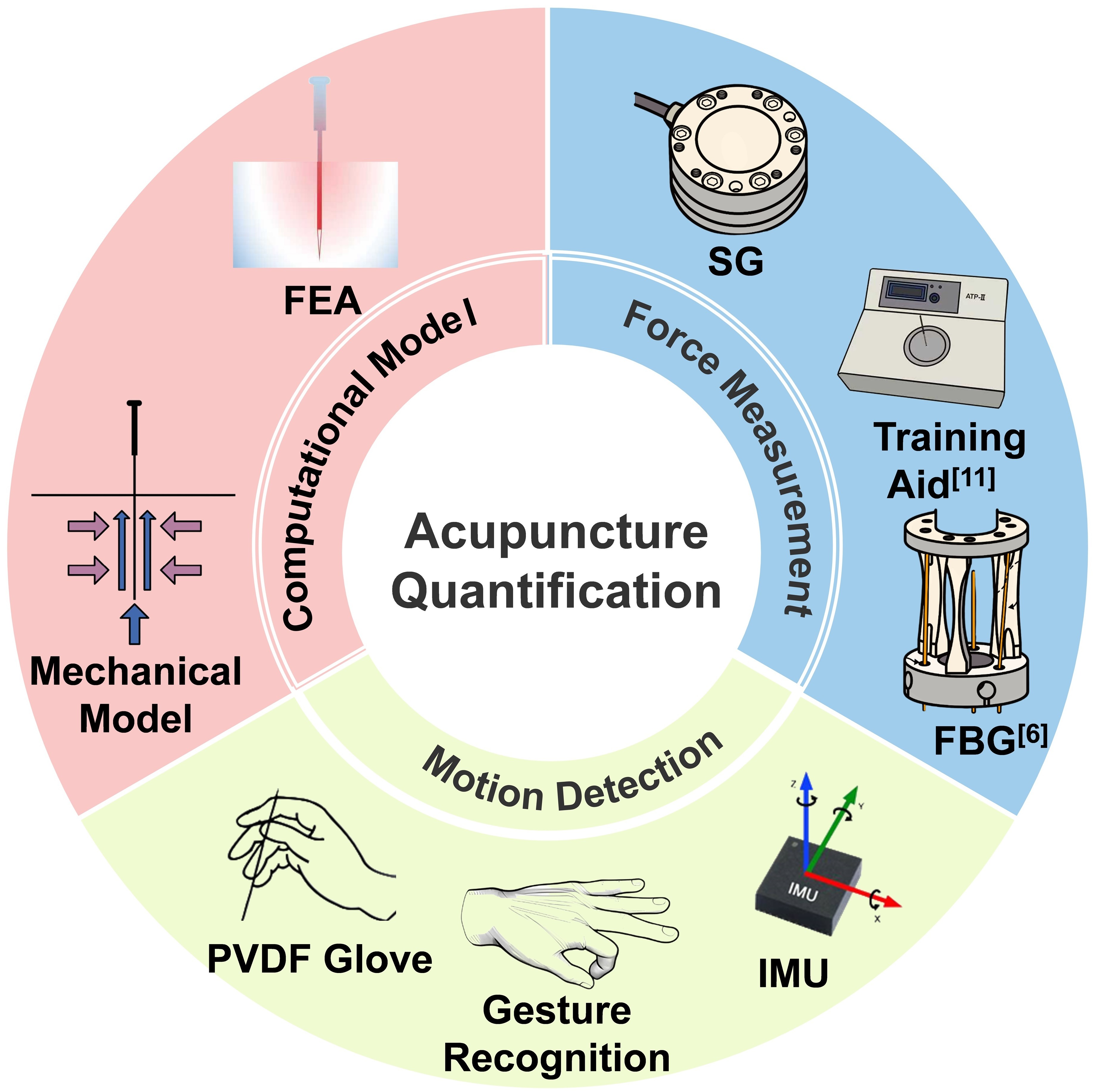}
\caption{Main methods of acupuncture quantification.\\
FEA: finite element analysis; SG: strain gauge; FBG: fiber Bragg grating; IMU: inertial measurement unit.
}
\label{fig_2}
\end{figure}

\begin{table*}[t]
\centering
\caption{Summary of the Methods-of-the-Art of Acupuncture Manipulation}
\begin{tabular}{@{}cccccccccc@{}}
\toprule
Research & \makecell{Force or \\Torque} & \makecell{Accuracy \\(\%F.S.)} & \makecell{Acceleration \\or Velocity} & Displacement & Angle & Sample Rate & \makecell{Installation \\Position} &Volume (cm\textsuperscript{3}) & Weight (g) \\
\midrule
{[10]}   & YES & 0.69\%  & NO  & NO  & NO  & 25Hz   & Needle                  & 3.63 & 3 \\
\addlinespace[1ex]
{[11]}   & YES & \textbackslash & YES & YES & YES & \textbackslash & Phantom                 & \textbackslash & \textbackslash \\
\addlinespace[1ex]
{[12]}   & YES & 0.5\%   & YES & NO  & NO  & 50Hz   & Needle                  & 3.29 & 9 \\
\addlinespace[1ex]
{[13]}   & YES & \textbackslash & NO  & YES & NO  & \textbackslash & Needle                  & 6    & \textbackslash \\
\addlinespace[1ex]
{[23]}   & YES & 0.67\%  & NO  & NO  & NO  & \textbackslash & Needle                  & 11   & \textbackslash \\
\addlinespace[1ex]
{[6]}    & YES & 3.5\%   & YES & NO  & YES & 100Hz  & \makecell{Needle +\\Under the Hand} & 1.5  & \makecell{0.7\\(only F/T sensor)} \\
\addlinespace[1ex]
{[25]} & YES & \textbackslash & YES & NO & NO & \textbackslash & \raisebox{-0.1\height}{Under the Hand} & \textbackslash & \textbackslash \\
\addlinespace[1ex]
{[24]} & YES & \textbackslash & NO & YES & NO & \textbackslash & \raisebox{-0.1\height}{Under the Hand} & \textbackslash & \textbackslash \\
\addlinespace[1ex]
\textbf{Our} & \textbf{YES} & \textbf{0.45\%} & \textbf{YES} & \textbf{YES} & \textbf{YES} & \textbf{100Hz} & \textbf{Needle} & \textbf{0.828} & \makecell{\textbf{1.3}\\\textbf{(all sensors+needle)}} \\
\bottomrule
\end{tabular}
\end{table*}

Acupuncture quantification methods primarily focus on three categories: computational mode, force measurement, and motion detection, as illustrated in Fig. 2.  Related research is listed in Table I. The following will provide a detailed introduction.
\subsection{Computational Model}
Acupuncture manipulation involves a complex mechanical process involving the insertion of a needle into the skin. Researchers have developed constitutive mechanical models to describe skin tissue behavior, including linear elastic [16], viscoelastic [17], and viscohyperelastic models [18]. Regardless of the specific model employed, the calculation of the reactive forces exerted by the tissue on the needle based on the velocity and displacement parameters. Accurate acquisition of these parameters is crucial for model validation. Yao et al. [19] established axial mechanical models for the needle, as well as models for the lifting-thrusting [20] and twirling-rotating operations\cite{ref21}, thus elucidating the key factors affecting the force exerted on the needle. They highlighted that the depth of needle insertion is closely correlated with therapeutic efficacy. Liang et al. [22] reviewed the mechanical effects of acupuncture, suggesting that variations in lifting-thrusting velocity and displacement can lead to changes in the force exerted by biological tissues on the needle. They emphasized that the mechanical parameters related to acupuncture directly influence the regulation of neural functions.
\subsection{Force Measurement}
Existing studies have focused on the development of high-precision acupuncture force sensors to accurately measure force variations during acupuncture. Xu et al. [10] developed an acupuncture force sensor with a novel elastomer design, incorporating hollow thin-walled cross beams and an inverted T floating beam structure, ensuring a coupling error of less than 2\%. Liang et al. [9] developed a three-dimensional fiber Bragg grating (FBG)  force sensor to detect needing force, with a measurement range of ±5 N and a resolution of 4.95 mN. Shi et al. [23] designed an FBG-based sensor featuring a force-sensitive flexure element with excellent linearity, achieving a measurement range of ±20 N and a resolution of 4.9 mN.
\subsection{Motion Detection}
Motion detection methods primarily utilize measurement technologies such as motion sensing and visual perception to detect the acupuncture movement parameters. Su et al. [24] developed a PVDF film-based array finger to capture the kinematic information of the physicians, achieving a test accuracy of 92.45\%  utilizing the CatBoost ensemble learning model as a classification approach. Gao et al. [25] designed a piezoelectric force sensing glove combined with Leap Motion gesture monitoring technology to extract key parameters, such as peak stress, average stress, and dispersion, to evaluate the acupuncture manipulation. Liang et al. [6] developed an FBG-FT-IMU acupuncture measurement system, which uses three fiber Bragg gratings (FBGs) to constitute a force/torque sensor and integrates with a gesture measurement glove equipped with an IMU to capture acupuncture motion information. The system performed classification tests on four basic acupuncture manipulations using a CNN-LSTM model, achieving a classification accuracy of 95.19\%.

\subsection{Remaining Issues}
Although the aforementioned methods have found wide application in the quantification of acupuncture manipulation,  further exploration is needed for the physical relationships among critical parameters as well as their direct measurement methods. Specifically, within model-based methods, the current acupuncture mechanical analysis model is a kinematic model. Three degrees of freedom within the dynamics of twirling and rotating operations have yet to be established. It is urgent to improve the dynamic model and its parameter detection. In force measurement methods, it is important to investigate how to further reduce the sensor's size and weight to enhance the portability of needle-holding operations. In motion detection methods, efforts should focus on directly measuring the needle's kinematic and dynamic parameters, rather than relying on indirectly deriving them through hand motion analysis.

Therefore, this study proposes a miniaturized sensing acupuncture needle and a novel acupuncture quantification system designed to quantify the complex motion information during acupuncture. By integrating kinematic and dynamic models, the system enables the interpretation of the physical meanings of key parameters. Through the FAVD-Acusensor and the acupuncture detection system, this framework enables the direct detection of critical parameters during acupuncture. The proposed framework aims to reveal the mechanical characteristics of different acupuncture manipulations, providing valuable tools and methods for acupuncture education, heritage preservation, and the elucidation of acupuncture mechanisms.
\section{SYSTEM METHODS AND MATERIALS }
This chapter establishes the kinematic and dynamic models of acupuncture, identifying the key parameters that need to be detected during the process. On this foundation, a miniaturized sensing acupuncture needle and an acupuncture quantification system are designed to address these critical parameters, achieving a digital quantitative description of acupuncture manipulations.

\subsection{Kinematic and Dynamic Models of Acupuncture }
\begin{figure}[t]
\centering
\includegraphics[width=\linewidth]{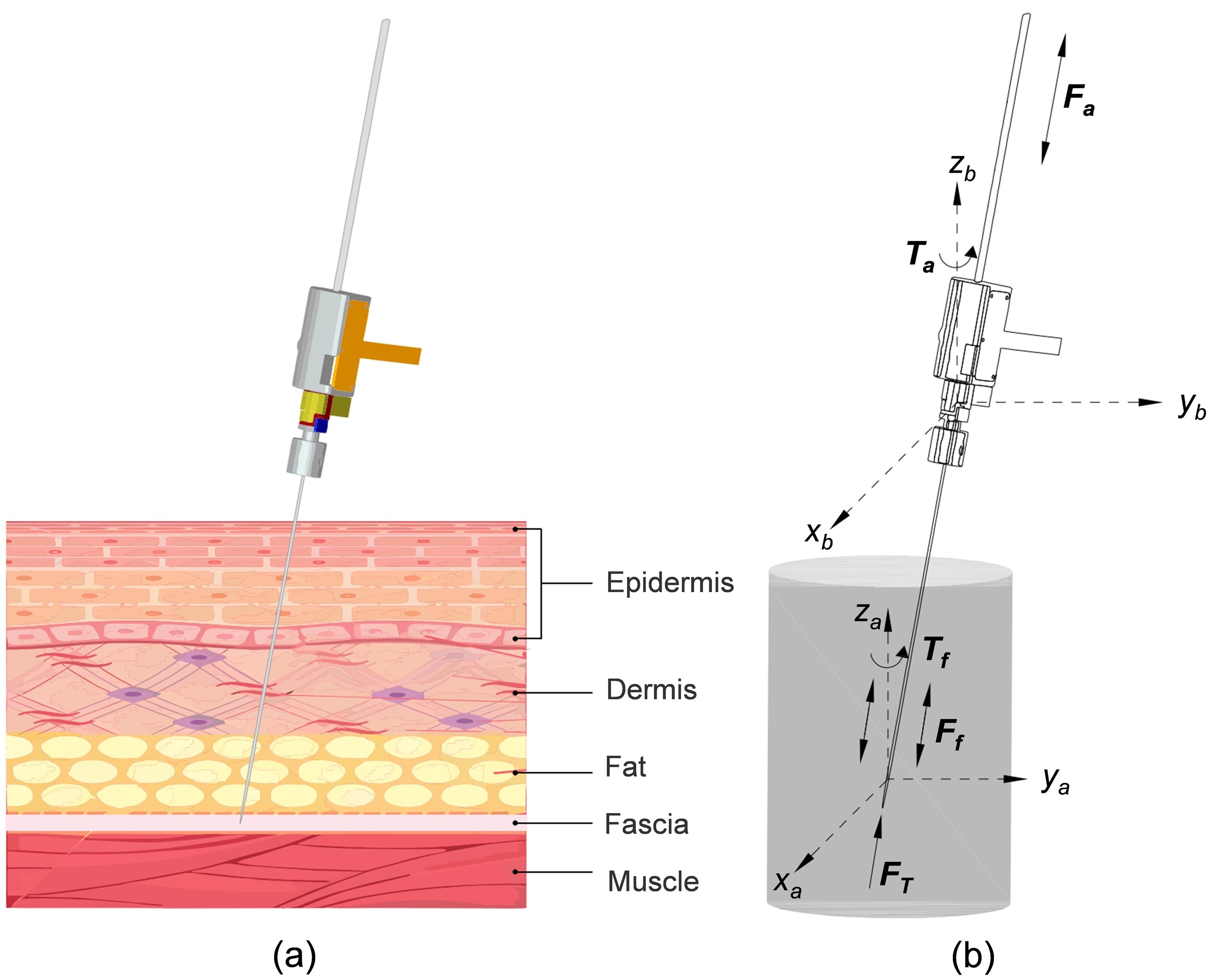}
\caption{(a) Diagram of needle-tissue interaction. (b) Force analysis of acupuncture needle.}
\label{fig_3}
\end{figure}
Acupuncture is a complex process characterized by the interaction between the needle and human tissues. Human tissues can be divided into multiple layers \cite{ref19, ref20}, as illustrated in Fig. 3(a). To describe the motion state of the needle, it is essential to analyze the forces acting upon it, as demonstrated in Fig. 3(b).

The kinematic and dynamic models of acupuncture are presented in Equations (1) and (2) , respectively, which describe the translational and rotational motion of the needle. 
\begin{equation}
\bm{m} \, \dot{\bm{v}} = \bm{F}_a - \bm{F}_T - \bm{F}_f 
\end{equation}
\begin{equation}
\bm{I}_{\bm{z}} \, \dot{\bm{\omega}} + \bm{\omega}^\times (\bm{I}_{\bm{z}} \bm{\omega}) = \bm{T}_a - \bm{T}_f 
\end{equation}

where $\bm{m}$ denotes the total mass of the sensing acupuncture needle. $\bm{v}$ denotes the velocity of motion, while $\dot{\bm{v}}$ denotes the acceleration of motion.  $\bm{F}_a$ is the applied force, $\bm{F}_T$ is the needle tip force, and $\bm{F}_f$ is the frictional force. $\bm{I}_{\bm{z}}$ is the inertia matrix of the sensing acupuncture needle, $\bm{\omega}$ is the angular velocity, $\dot{\bm{\omega}}$ is the angular acceleration, and $\bm{\omega}^\times$ is the angular velocity cross product matrix. $\bm{T}_a$ is the applied torque, while $\bm{T}_f$ is the frictional torque. 

The kinematics model of the needle characterizes its lifting-thrusting operations, whereas the dynamics model describes the twirling-rotating operations. As shown in Equation (1), when the needle is undergoing translation, it is subject to the applied force, the needle tip force, and the frictional force acting together. The applied force is the force exerted by the physicians through their fingers onto the needle. The needle tip force is the compressive force between the needle tip and the contacted tissue during the needle insertion process. The frictional force is the friction generated between the needle and all the pierced tissues during the insertion process. The calculation methods for the needle tip force and frictional force can refer to the constitutive models of skin tissue\cite{ref19, ref20}. In these models, both the needle tip force and frictional force are directly related to velocity and displacement, emphasizing the importance of detecting velocity and displacement.

The rotational motion of the needle can be described using the dynamic model as in Equation (2), with the needle subject to both the applied torque and the frictional torque during twirling-rotating operations. The method for calculating frictional torque is detailed in Equation (3).
\begin{equation}
\bm{T}_f = \sum_{i=1}^{n} \left( \mu_i \bm{F}_{i} r + \bm{\eta}_i \, \bm{\omega} r^2 \bm{L}_i \right)
\end{equation}
where, $\mu_i$ denotes the friction coefficient between the needle and the \textit{i-th} tissue layer, $\bm{F}_{i}$ denotes the normal force exerted by the \textit{i-th} tissue on the needle, $r$ is the radius of the needle, $\bm{\eta}_i$ is the viscosity coefficient of the \textit{i-th} tissue, and $\bm{L}_i$ is the contact length between the needle and the \textit{i-th} tissue layer. 

In this study, the twirling-rotating angular velocity is utilized to reflect rotational motion under the resultant torque, with the effect of frictional torque considered. This approach comprehensively assesses the mechanical characteristics of twirling-rotating operations, simplifying both the size and weight of the sensing needle while fulfilling the research objective of quantifying acupuncture. In contrast, detecting applied torque through torque sensors fails to accurately capture the dynamic characteristics of the operation \cite{ref12, ref26}.  Due to the influence of frictional torque, measuring only the applied torque does not fully capture the rotational motion under the resultant torque.

\subsection{Design of Acupuncture Quantification System}

\begin{figure*}[t]
\centering
\includegraphics[width=0.9\linewidth]{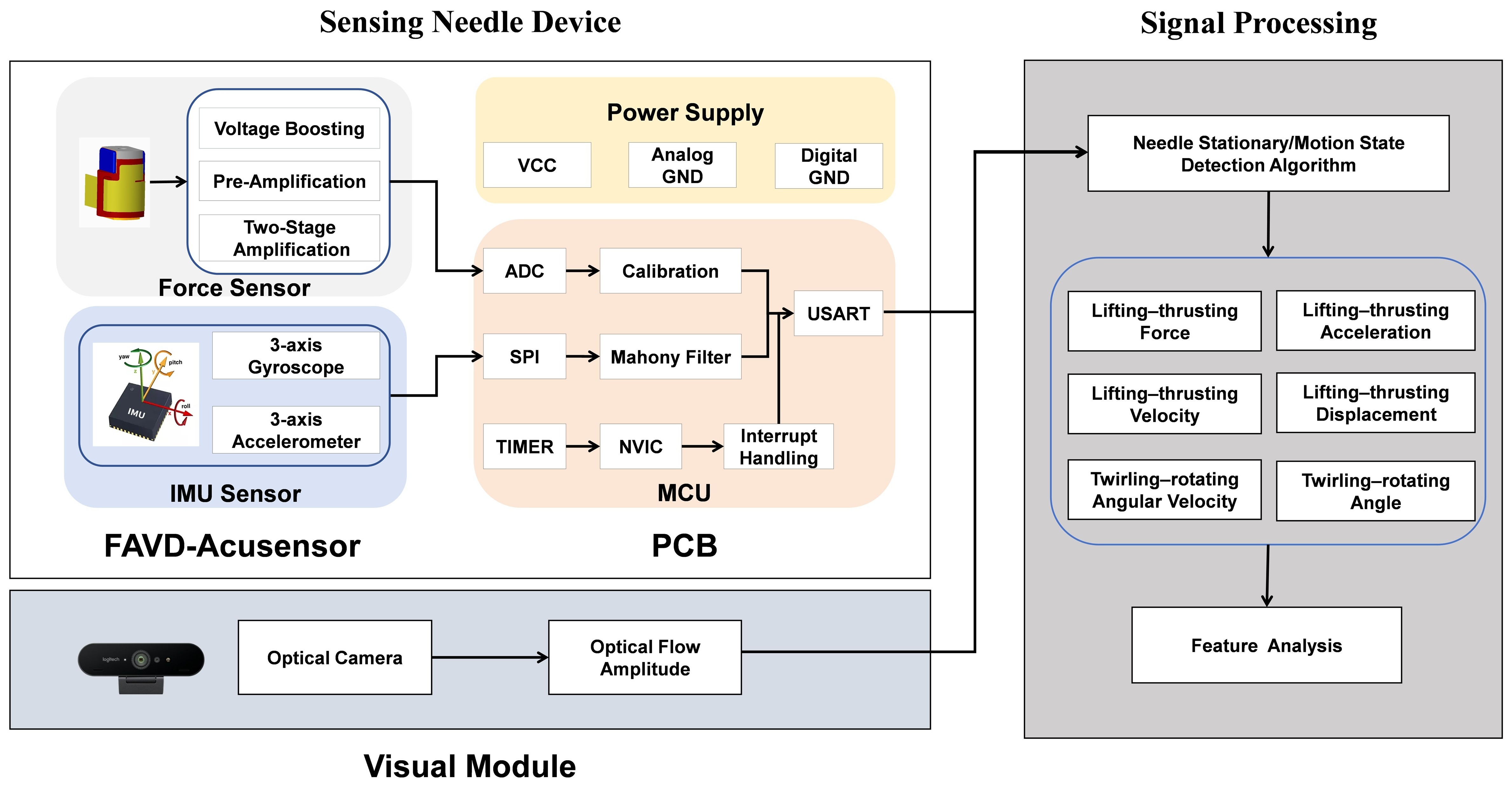}
\caption{System framework of the proposed acupuncture quantitative system, including the sensing needle device, visual module, and signal processing processes.}
\label{fig_4}
\end{figure*}

\begin{table}[t]
\centering
\caption{System Specifications}
\begin{tabular}{@{}cc@{}}
\toprule
Parameter & Value \\
\midrule
Overall Dimensions & 0.828 cm\textsuperscript{3} \\
\addlinespace[1ex]
Weight & 1.3 g \\
\addlinespace[1ex]
Physiological Measurements & \makecell{Force, Acceleration,  \\Velocity, Displacement, \\Angular velocity, and Angle,} \\
\addlinespace[1ex]
Sensor Sample Rate & 100 Hz \\
\addlinespace[1ex]
Nonlinearity error of force sensor & 0.45\% \\
\addlinespace[1ex]
Gyroscope Angle Random Walk & $\pm 4$ mdps$/\sqrt{\text{Hz}}$ \\
\addlinespace[1ex]
Accelerometer Velocity Random Walk & 100 $\mu$g$/\sqrt{\text{Hz}}$ \\
\addlinespace[1ex]
Gyroscope Sensitivity Scale Factor & 16.4 LSB/(dps) \\
\addlinespace[1ex]
Accelerometer Sensitivity Scale Factor & 8192 LSB/g \\
\bottomrule
\end{tabular}
\end{table}

\subsubsection{System Framework}

From the dynamic analysis, it is evident that the critical parameters include lifting-thrusting force, acceleration, velocity, displacement, as well as twirling-rotating angular velocity and angle. To achieve the measurement of these parameters, the FAVD-Acusensor and the acupuncture quantification system were proposed. The hardware components of the acupuncture quantification system include the sensing needle device and the visual module, as shown in Fig. 4. Among them, the sensing needle device is divided into two parts: the FAVD-Acusensor and the printed circuit board (PCB).

The FAVD-Acusensor integrates a tension and compression force sensor and an IMU; the PCB incorporates force signal processing circuits, power management circuits, and a microcontroller unit (MCU). The visual module, in conjunction with the IMU, calculates the segmented intervals of the needle insertion state, which are utilized for the purpose of segmented integration in determining the lifting-thrusting velocity and displacement. The technical specifications are summarized in Table II, demonstrating the device's capability for high-quality recordings.

\subsubsection{FAVD-Acusensor Design}

\begin{figure}[t]
\centering
\includegraphics[width=\linewidth]{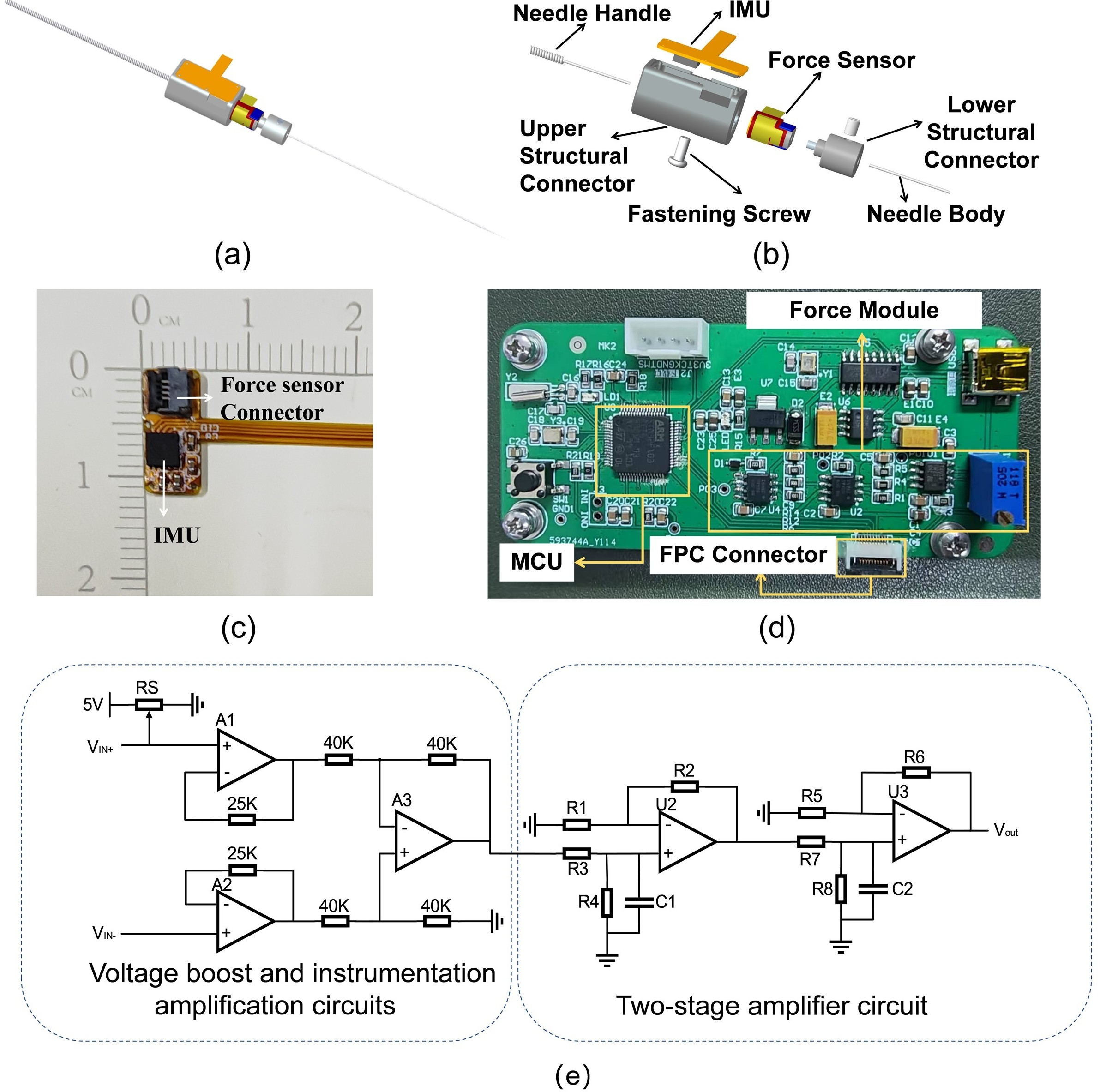}
\caption{(a) Design diagram of the FAVD-Acusensor. (b) Layered structure of the FAVD-Acusensor. (c) Physical diagram of the IMU sensor. (d) Device hardware diagram. (e) Key circuit for force signal processing.}
\label{fig_5}
\end{figure}

This section details the design of the FAVD-Acusensor integrating force and IMU sensors, specifically tailored for the kinematic and dynamic parameters of acupuncture procedures. The overall structure of the FAVD-Acusensor is shown in Fig. 5(a). QLA414 (FUTEK) is selected as the force sensor, characterized by its compact size (4 × 5 mm) and lightweight (0.5 g), with a range of 0-22.2 N.

To implement the detection of lifting-thrusting force, the design employs upper and lower structural connectors to achieve rigid connections between the needle handle, the IMU, the force sensor, and the needle body, as illustrated in Fig. 5(b).

The IMU used in this study is the ICM20602 (InvenSense), which integrates a three-axis accelerometer and a three-axis gyroscope. To enhance integration with the force sensor, a flexible printed circuit (FPC) was designed with dimensions of 5.5 × 11.0 mm, offering a more compact solution compared to commercial IMU modules, as illustrated in Fig. 5(c). The quaternion method is utilized to describe the angles of the needle. In particular, considering the vertical operational characteristics of the needle insertion procedure and the perpendicular installation of the IMU, the roll angle is employed to characterize the twirling-rotating angle during acupuncture procedures.

Methods for calculating twirling-rotating angle include complementary filtering [27], improved complementary filtering [28], and advanced variants of Kalman filtering\cite{ref29, ref30}. In this study, complementary filtering algorithms are implemented for accurate attitude estimation. This approach ensures high precision while demanding minimal computational resources.
\subsubsection{Printed Circuit Board (PCB) and Design Rationale}
A photograph of the PCB is shown in Fig. 5(d), including a force signal processing circuit, a power management circuit, and an MCU regulation circuit. In this study, we utilize the STM32F103RCT6 microcontroller (STMicroelectronics), featuring a 12-bit analog-to-digital converter (ADC) with a sampling rate of 100 Hz. 

The block diagram detailing the design of the force signal processing circuit is shown in Fig. 5(e). The force sensor has an output sensitivity of 0.5 mV/V and is powered by a 5V excitation voltage. Due to the weak nature of the output voltage signal and the presence of common-mode noise, 1/f noise, and thermal noise after the bridge conversion,  dedicated amplification circuits are required to optimize signal quality. This study proposes a three-stage amplification approach, with the first stage also designed for voltage lifting.

The design of the instrumentation amplification circuit requires a high common-mode rejection ratio (CMRR) and high input impedance. In this design, the IN- terminal of the pressure signal undergoes voltage boosting through a biasing circuit, elevating the reference voltage to 1.65 V. The system employs the instrumentation amplifier INA128 (Texas Instruments) for pre-amplification, achieving a signal amplification factor of 6 times.

The last two stages are constructed using the OPA604 operational amplifier (Texas Instruments). The signal, which has undergone differential amplification by a instrumentation amplifier, is inputted to the non-inverting terminal of the first-stage OPA604. Both stages are configured to achieve a gain of 10. In conjunction with the 6 times gain from the pre-instrumentation amplification stage, the system collectively realizes a total voltage gain of 600 times.
\subsubsection{Intelligent Vision Module}
\begin{figure}[t]
\centering
\includegraphics[width=0.95\linewidth]{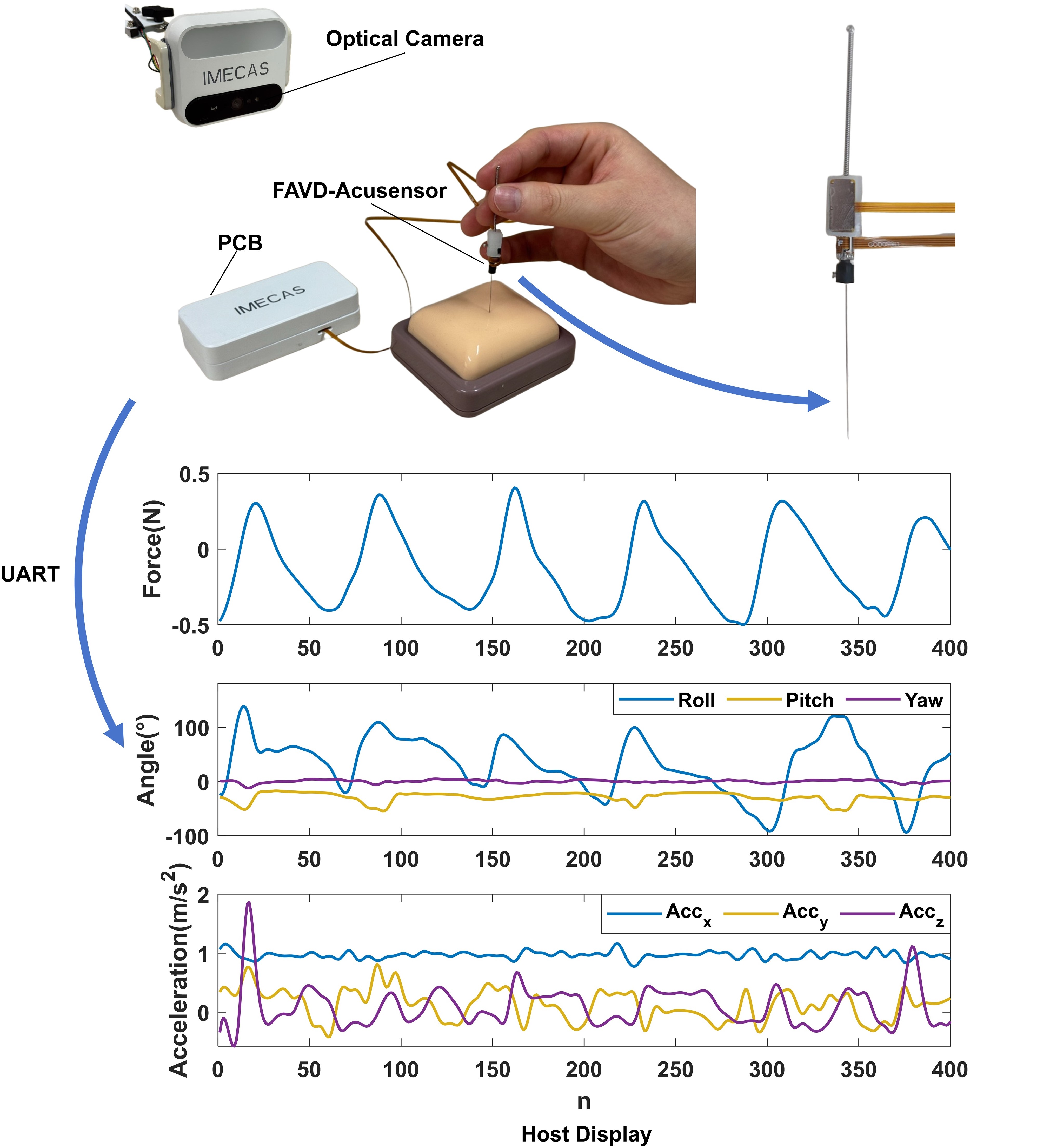}
\caption{Physical diagram of the acupuncture quantitative system.}
\label{fig_6}
\end{figure}
This study employs visual technology to achieve digital recording of acupuncture and assessment of stationary/motion states. Logitech C1000e camera (1920 $\times$ 1080 @30fps) is utilized as the acquisition device. To optimize imaging quality, this module integrates an adjustable LED illumination module and employs a diffused reflection design, significantly reducing glare from the metal needle and ultimately achieving high dynamic range imaging of the acupuncture.The fusion technology of visual and IMU data provides decision support for the calculation of the lifting-thrusting velocity and displacement. The video frame rate is 30 Hz, while the IMU sampling rate is 100 Hz. Since needle insertion is a continuous motion, interpolation is meaningful in this context. Cubic spline interpolation is then applied to align the visual data with the IMU data. The effectiveness of the interpolation is validated through experiments using a high-speed camera (Model: Insta360 X5, 3840 $\times$ 1920 @100 fps) to ensure the quality and reliability of data. In Section III C(2), a detailed introduction to the fusion of visual and IMU data for accurate recognition of needle states will be provided. Fig. 6 illustrates the physical diagram of the acupuncture quantification system, including the sensing needle device and the visual module.

\subsection{Design of multimodal signal processing chains}
\begin{figure*}[t]
\centering
\includegraphics[width=\linewidth]{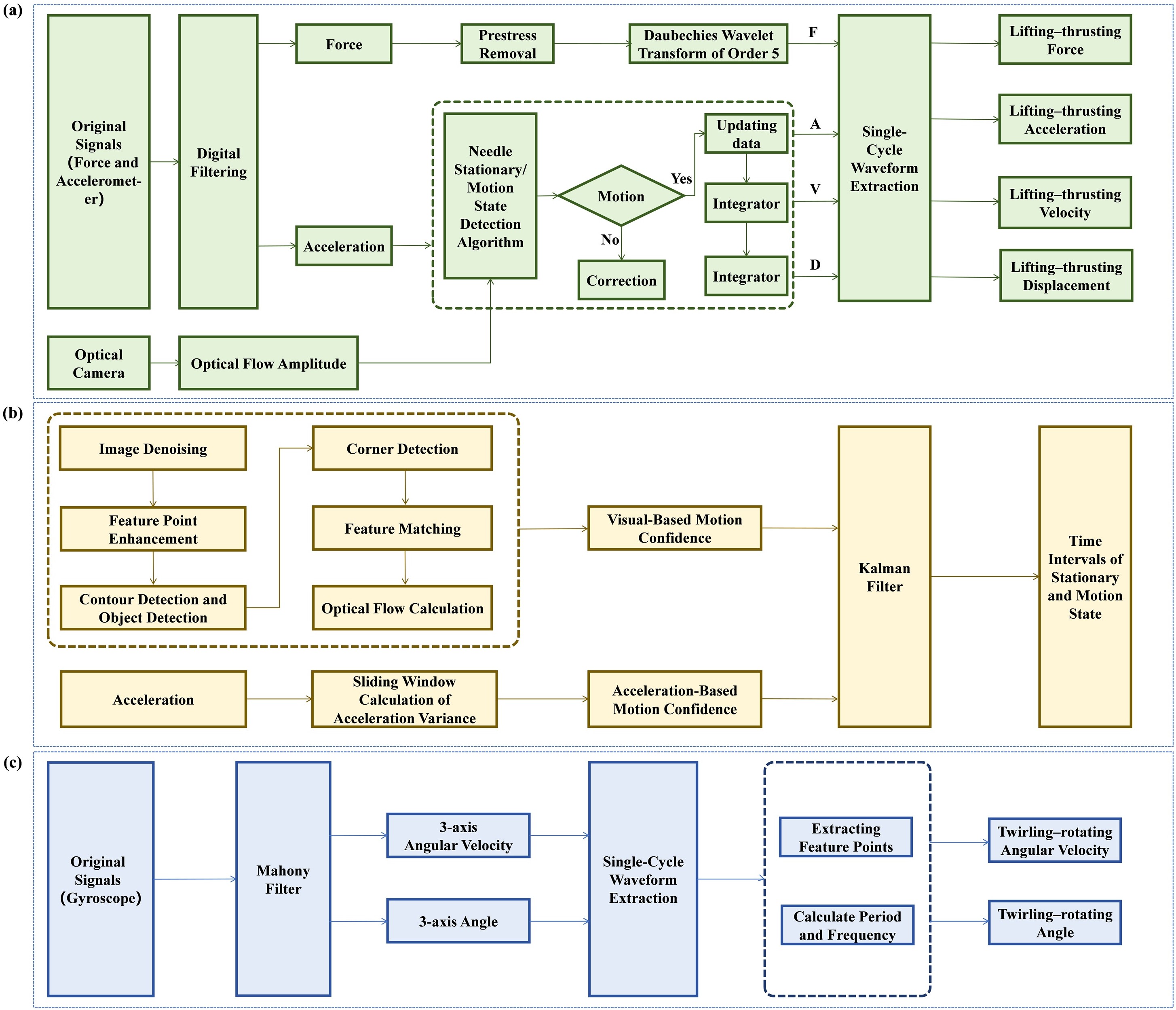}
\caption{(a) Force, acceleration, velocity, and displacement signal processing in lifting-thrusting operation. (b) Needle Stationary/Motion State Detection Algorithm. (c) Angular velocity and angle signal processing in twirling-rotating operation. }
\label{fig_7}
\end{figure*}

\subsubsection{Digital Filtering}
Based on Equation (1), the lifting-thrusting force and acceleration signals of the needle provide an objective description of the lifting-thrusting process. The raw signals from the force sensor undergo prestress measurement and compensation, 50 Hz notch filtering, and Daubechies wavelet transform of order 5 to obtain the filtered lifting-thrusting force. The raw signals from the accelerometer are processed through a digital low-pass filter within the chip (bandwidth of 44.8 Hz) to obtain the filtered acceleration signals, as illustrated in Fig. 7(a). Based on Equation (2), the angular velocity of the needle's rotation provides an objective description of the twirling-rotating operation. The raw signals from the gyroscope undergo Mahony filter to obtain the angular velocity and angles, as illustrated in Fig. 7(c).

\subsubsection{Calculation of lifting-thrusting velocity and Displacement}

During acupuncture needle insertion, the accuracy of velocity and displacement calculations is affected by errors in acceleration and velocity signals. By integrating visual and IMU data, the four stages of a single-cycle lifting-thrusting operation are identified. The lifting-thrusting velocity and displacement are calculated using segmented integration, thereby reducing cumulative errors and enhancing computational accuracy. The calculations of velocity and displacement are performed after data collection.

The single-cycle thrusting-lifting operation can be subdivided into four distinct stages, as described below:

Stage 1 ($S_1$): The thrusting stage, during which the needle is downward inserted.

Stage 2 ($S_2$): The interval stage between the end of thrusting and the start of lifting.

Stage 3 ($S_3$): The lifting stage, during which the needle is upward lifted.

Stage 4 ($S_4$): The interval stage between the end of lifting and the start of the next cycle's thrusting stage.

During the intervals ($S_2$ and $S_4$), there is no active acupuncture operation, and the needle displacement remains nearly constant. However, due to the inability of the hand holding the needle to maintain an absolutely stationary state during operation, the acceleration signal does not equal zero. If acceleration signals during the intervals are directly integrated, it will lead to errors in the calculation of the lifting-thrusting velocity and displacement, thereby affecting the overall computational accuracy. This paper employs the Kalman filter-based confidence fusion method for detecting the static/motion state of acupuncture, integrating IMU and visual data to achieve accurate identification of the acupuncture state, as illustrated in Fig. 7(b).

The Kalman filter is suitable for continuous state estimation. Motion confidence is defined as a state vector, with confidence values ranging from [0,1]. A higher value indicates a higher probability of motion. The state vector is defined as follows:
\begin{equation}
x_k = p_k
\end{equation}
where $p_k$ represents the motion confidence at the current moment.

The prior estimate equation is denoted as: 
\begin{equation}
x_{k|k-1} = x_{k-1} + w_k
\end{equation}
where $w_k \sim N(0, Q)$.  $w_k$ is the process noise and $Q$ is the covariance matrix.

The measurement vector is:
\begin{equation}
\bm{z}_k =
\begin{bmatrix}
1 - \exp\left(-\dfrac{z_{\text{acc}}}{\tau_{\text{acc}}} \right) \\
1 - \exp\left(-\dfrac{z_{\text{vis}}}{\tau_{\text{vis}}} \right)
\end{bmatrix}
\end{equation}
where, $z_{\text{acc}}$ represents the variance of the acceleration signal, which is obtained by calculating the variance of the acceleration signal within a sliding window. $\tau_{\text{acc}}$ denotes the acceleration-based motion threshold. $z_{\text{vis}}$ represents the optical flow magnitude, which is calculated as follows: the target detection model is utilized to locate the outline of the needle, and the coordinates of the center point of the needle handle are extracted. The displacement vectors for each pixel between two consecutive video frames are computed via the optical flow method, and the corresponding optical flow magnitude is derived. Subsequently, the optical flow algorithm is applied to compute the displacement vector between consecutive frames, and after filtering, the optical flow magnitude is obtained. $\tau_{\text{vis}}$ denotes the optical flow magnitude threshold.

The measurement equation is:
\begin{equation}
\bm{z}_k = \bm{H} \bm{x}_k + \bm{v}_k
\end{equation}
where the measurement matrix $\bm{H} = \begin{bmatrix} 1 \\ 1\end{bmatrix}$, $\bm{v}_k \sim \bm{N}(\bm{0}, \bm{R})$. The noise covariance matrix $\bm{R}$ represents the sensor reliability. After establishing the estimation and measurement equations, the time series of motion confidence can be obtained through the time-update step and measurement-update step of the Kalman Filter [31]. A threshold is established to distinguish between motion and stationary states. When the motion confidence exceeds this threshold, the state is categorized as ``motion"; otherwise, it is designated as ``stationary." Consequently, the corresponding time intervals for both states can be identified.

On this basis, a segmented integration method is employed to calculate the lifting-thrusting velocity and displacement. The calculation process for velocity and displacement is illustrated in Fig. 7(a). Based on the needle stationary/motion state detection algorithm, time intervals of the states are obtained. If the needle is in the motion state, the acceleration data is updated and the lifting-thrusting velocity and displacement are calculated. If the needle is in the stationary state, the acceleration signal needs to be corrected by setting the acceleration to zero, maintaining the velocity at zero, and keeping the displacement value constant. The formula for calculating the lifting-thrusting velocity is as follows.
\begin{equation}
v(n) =
\left\{
\begin{aligned}
&v(n\!-\!1) \!+\! \dfrac{a(n) \!+\! a(n\!-\!1)}{2}  \Delta t && n \in \{S_1, S_3\} \\
&0 && n \in \{S_2, S_4\}
\end{aligned}
\right.
\end{equation}
where $a(n)$ represents the resultant acceleration at the \textit{n-th} sampling point, $\Delta t$ represents the sampling interval, $n \in \{S_1, S_3\}$ indicates the current state is in the motion state, and $n \in \{S_2, S_4\}$ indicates the current state is in the stationary state. 

The calculation formula for lifting-thrusting displacement is:
\begin{equation}
d(n) =
\left\{
\begin{aligned}
&d(n\!-\!1) \!+\! \dfrac{\mathbf{v}(n) \!+\! \mathbf{v}(n\!-\!1)}{2} \Delta t && n \in \{S_1, S_3\} \\
&d(n\!-\!1) && n \in \{S_2, S_4\}
\end{aligned}
\right.
\end{equation}
where $d(n)$ and $v(n)$ represent the displacement and the resultant velocity  at the \textit{n-th} sampling point, respectively.
 
By detecting the needle stationary/motion states and then applying a segmented integration method, the noise interference in the intervals ($S_2$ and $S_4$) can be effectively suppressed. This reduces the cumulative errors caused by noise for velocity and displacement calculations, thereby improving the accuracy of the results.
\subsubsection{Feature Extraction and Cycle Segmentation in Lifting-Thrusting Operation}
The method for defining the cycle of acupuncture manipulation is currently determined by the signal characteristics of the applied force. However, relying solely on the polarity and the rate of change of the applied force to define the cycle has its limitations\cite{ref32, ref33}. This is because acupuncture manipulation, in addition to being influenced by the applied force, is also affected by the combined effects of the needle tip force and frictional force, and factors such as force signal filtering all introduce errors. Therefore, to more accurately segment the lifting-thrusting operation cycle, one can integrate the information from lifting-thrusting displacement and acceleration signals to comprehensively determine the thrusting and lifting stages during the acupuncture process. This approach can overcome the limitations of using only force signals for cycle segmentation.

This study employs a methodology that integrates visual and IMU data to achieve the segmentation of the four stages in a single-cycle lifting-thrusting operation. Visual images contain timestamp information, which corresponds to the timestamp information in the force sensor and acceleration signals. Specifically, the single cycle of the acceleration signal is divided into four characteristic points: Thrusting Start, Thrusting End, Lifting Start, and Lifting End. These points are defined as follows. Thrusting Start (TS): the start point of the thrusting stage; Thrusting End (TE): the end point of the thrusting stage; Lifting Start (LS): the start point of the lifting stage; Lifting End (LE): the end point of the lifting stage. Peak detection and forward/backward search are implemented on the acceleration signal to determine the start and end points of the lifting and thrusting stages. This approach enables a comprehensive joint analysis of multiple features. Then, the lifting-thrusting operation cycles can be segmented based on the four characteristic points.

\subsubsection{Feature Extraction and Cycle Segmentation in Twirling-Rotating Operation}

Compared to lifting-thrusting operations, the interval time between twirling-rotating operations is extremely brief and is no longer subdivided into intervals. The twirling-rotating operation consists of two distinct stages: left-ward twirling stage and right-ward twirling stage. During the left-ward twirling stage, the needle rotates clockwise, corresponding to a negative angular velocity, with the angle progressively decreasing. This stage spans from the initiation of left-ward twirling to the onset of right-ward twirling. During the right-ward twirling stage, the needle rotates counterclockwise, corresponding to a positive angular velocity, with the angle progressively increasing. This stage extends from the initiation of right-ward twirling to the next left-ward twirling's start point. We define two characteristic points for twirling-rotating operation: the left twirling start point, which represents the initial point of left-ward twirling stage; and the right twirling start point, which represents the initial point of right-ward twirling stage. We utilize the "findpeaks" function to extract the peak and valley values from the angular waveform. Subsequently, the twirling-rotating operation cycles can be segmented by analyzing the two characteristic points within each cycle. Moreover, the twirling-rotating frequency can be derived by calculating the length of twirling-rotating operation cycles.
    
\section{TEST RESULTS AND ANALYSIS}

To validate that the FAVD-Acusensor is capable of accurately measuring kinematic and dynamic parameters, two types of experiments were conducted: sensor calibration experiments and acupuncture manipulation data collection experiments. In the data collection experiments, to verify the universal applicability of this tool across different skill levels and operational habits, and to analyze the operational characteristics of basic acupuncture manipulations, we selected five physicians from the Acupuncture and Moxibustion Hospital of China Academy of Chinese Medical Sciences, and Beijing Hospital of Traditional Chinese Medicine,Capital Medical University, who individually performed four basic acupuncture manipulations, including lifting-thrusting reinforcement (LTRF), lifting-thrusting reduction (LTRD), twirling-rotating reinforcement (TRRF), and twirling-rotating reduction (TRRD). The participants included three physicians with over 20 years of experience, one with over 10 years, and one with less than 10 years. Fresh pork was used as an experimental subject, and each physician collected 5 minutes of data under each basic acupuncture manipulation. During the experiment, the operator holds the FAVD-Acusensor, the visual module records the process in real time, and the collected data is transmitted to the host computer via USB.

\subsection{Calibration Experiments and Analysis}

\begin{figure}[t]
\centering
\includegraphics[width=\linewidth]{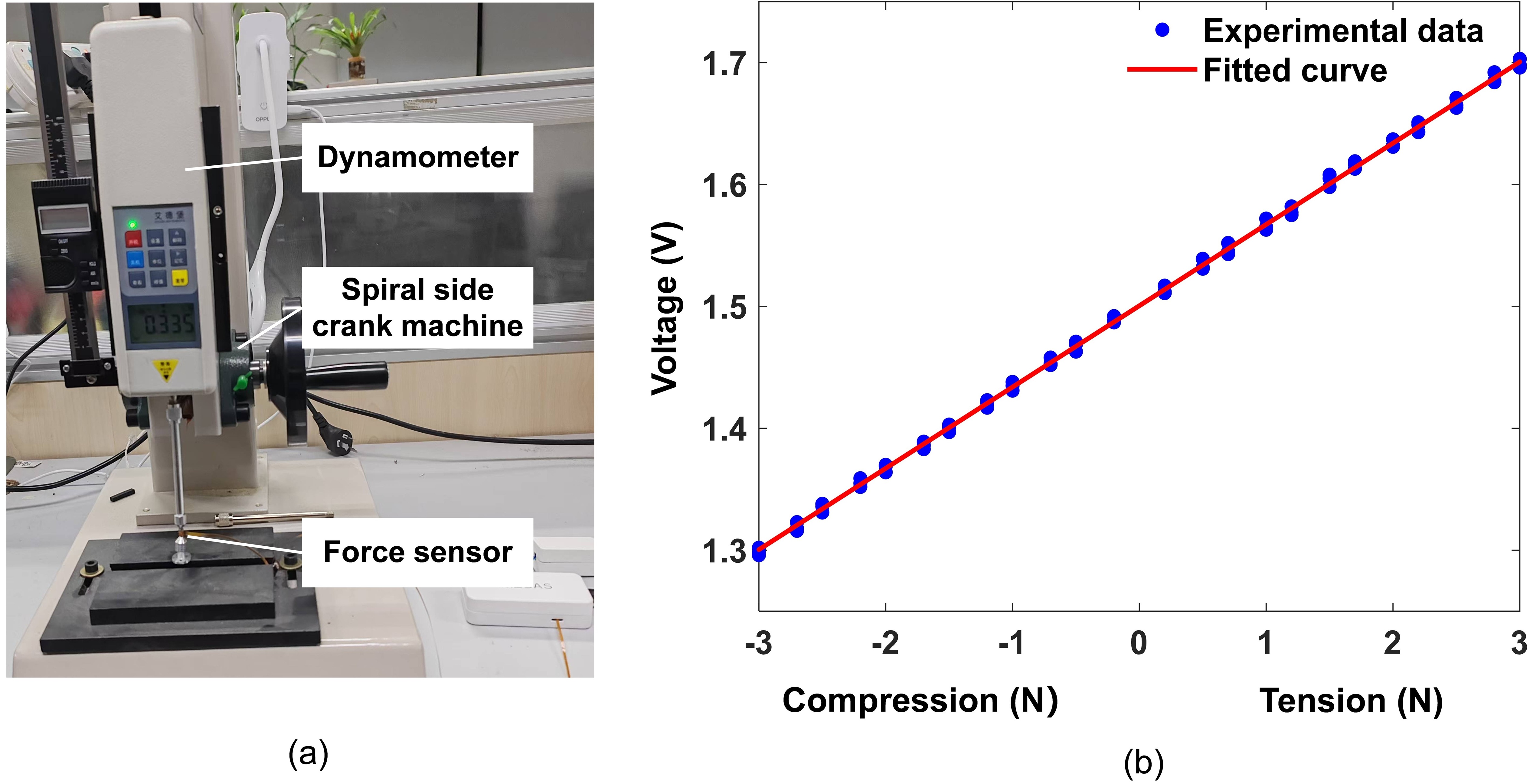}
\caption{Testing of force sensor. (a) Experimental setup for force calibration. (b) Output voltages of the force sensor with increased and decreased external force in the tension range of 0-3 N and compression range of -3-0 N. }
\label{fig_8}
\end{figure}

\subsubsection{Sensor Calibration}

The calibration experimental setup is illustrated in Fig. 8(a), where the dynamometer and force sensor are rigidly connected and fixed via a connecting fixture. Based on the principle of action-reaction forces having equal magnitude, the magnitude of the force applied to the force sensor can be determined. Subsequently, the relationship between different tension/compression force magnitudes and the output voltage of the force sensor was established. To perform the calibration, the force sensor amplification module was set to a magnification factor of 600 times, and calibration experiments were conducted three times, resulting in the applied force-relative voltage response curves shown in Fig. 8(b). These curves exhibit good linearity. The measurement precision of the force sensor was evaluated using the nonlinearity error, which was determined to be 0.45\%.

\subsubsection{IMU Error Analysis}

To evaluate the random errors and performance stability of the IMU, Allan variance [34] is commonly employed. Initially, the IMU is sampled for 2 hours, and the accelerometer and gyroscope data are fed into the Allan variance calculation function. A log-log plot of time intervals versus Allan variance is generated and plotted. Curve fitting is performed to determine the noise coefficients of random errors for the accelerometer and gyroscope, as shown in Fig. 9. In this study, we primarily consider bias instability, angle random walk, and velocity random walk. Through the analysis of random errors in the MEMS sensor, it was concluded that the sensor satisfies the requirements for calculating acupuncture manipulation information. The Allan variance fitting parameters are presented in Table III.

\begin{figure}[t]
\centering
\includegraphics[width=\linewidth]{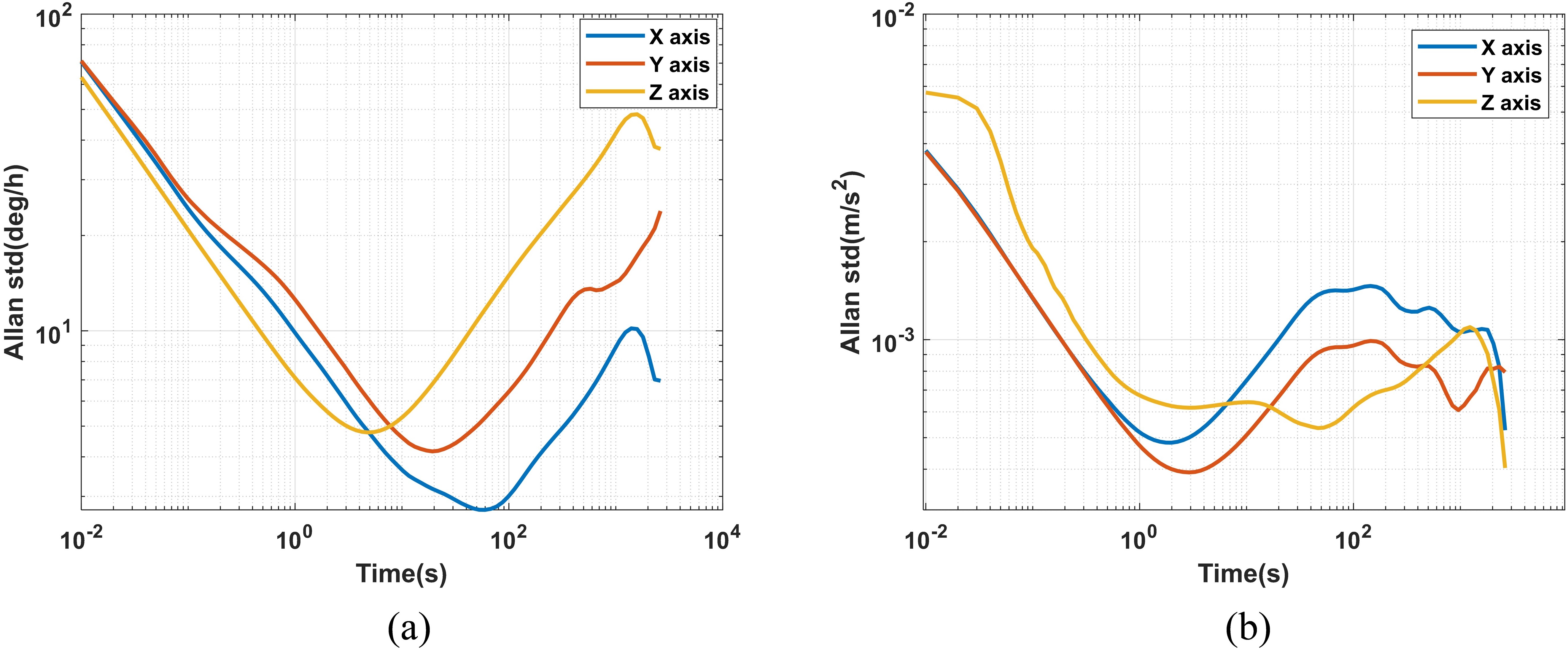}
\caption{Testing of the IMU sensor. (a) MEMS accelerometer Allan standard deviation and log-log plot of sampling time. (b) MEMS gyroscope Allan standard deviation and log-log plot of sampling time.}
\label{fig_9}
\end{figure}

\begin{table}[t]
\centering
\caption{Allan Variance Fitting Parameters}
\begin{tabular}{@{}cccccc@{}}
\toprule
& Parameter & Unit & X axis & Y axis & Z axis \\
\midrule
\multirow{2}{*}{Accelerometer} 
  & \makecell{Velocity \\Random Walk} & $\mu\text{g}/\sqrt{\text{Hz}}$ & 19.7 & 28.1 & 73.2 \\
  & \makecell{Bias \\Instability} & mg & 0.48 & 0.39 & 0.61 \\
\midrule
\multirow{2}{*}{Gyroscope} 
  & \makecell{Angle \\Random Walk} &$^\circ \cdot \text{h}^{-\frac{1}{2}}$  & 0.14 & 0.14 & 0.13 \\
  & \makecell{Bias \\Instability} & $ ^\circ \cdot \text{h}^{-1}$ & 2.72 & 4.16 & 4.78 \\
\bottomrule
\end{tabular}
\end{table}

\subsection{Displacement Test}

To measure the displacement of the FAVD-Acusensor, a height gauge was employed to record the displacement during the experiment. Specifically, the vernier caliper within the height gauge was adjusted to induce upward and downward movements of the FAVD-Acusensor. A total of 100 seconds of testing was conducted. Fig. 10(a) illustrates the experimental setup for the displacement test, while Fig. 10(b) provides a comparison between the measured displacement and the calculated displacement over a 10-second interval. The calculated displacement was obtained by employing the needle stationary/motion state detection algorithm and the corresponding displacement calculation procedure detailed in Section III C(2). The measured displacement was recorded by the height gauge in the video images, with cubic spline interpolation applied to process the data. The root mean square error (RMSE) is utilized as the evaluation metric for displacement accuracy, yielding a result of 1.2 mm, whereas the RMSE obtained by directly performing a double integration of the acceleration signal was 11.9 mm. Displacement estimation utilizing the integration of visual and IMU data offers significant advantages over methods relying solely on accelerometer data. By mitigating cumulative errors arising from acceleration integration, the fusion approach provides displacement estimates with higher accuracy.

\subsection{Feature Analysis of Acupuncture Manipulations}

This section discusses the characteristics of lifting-thrusting and twirling-rotating operations. From the experimental data, 100 complete cycles each for LTRF, LTRD, TRRF, and TRRD manipulations were extracted, totaling 400 acupuncture manipulation cycles. A single lifting-thrusting operation cycle was subdivided into thrusting and lifting stages, while a single twirling-rotating operation cycle was divided into left-ward twirling and right-ward twirling stages for detailed analysis.

\begin{figure}[t]
\centering
\includegraphics[width=\linewidth]{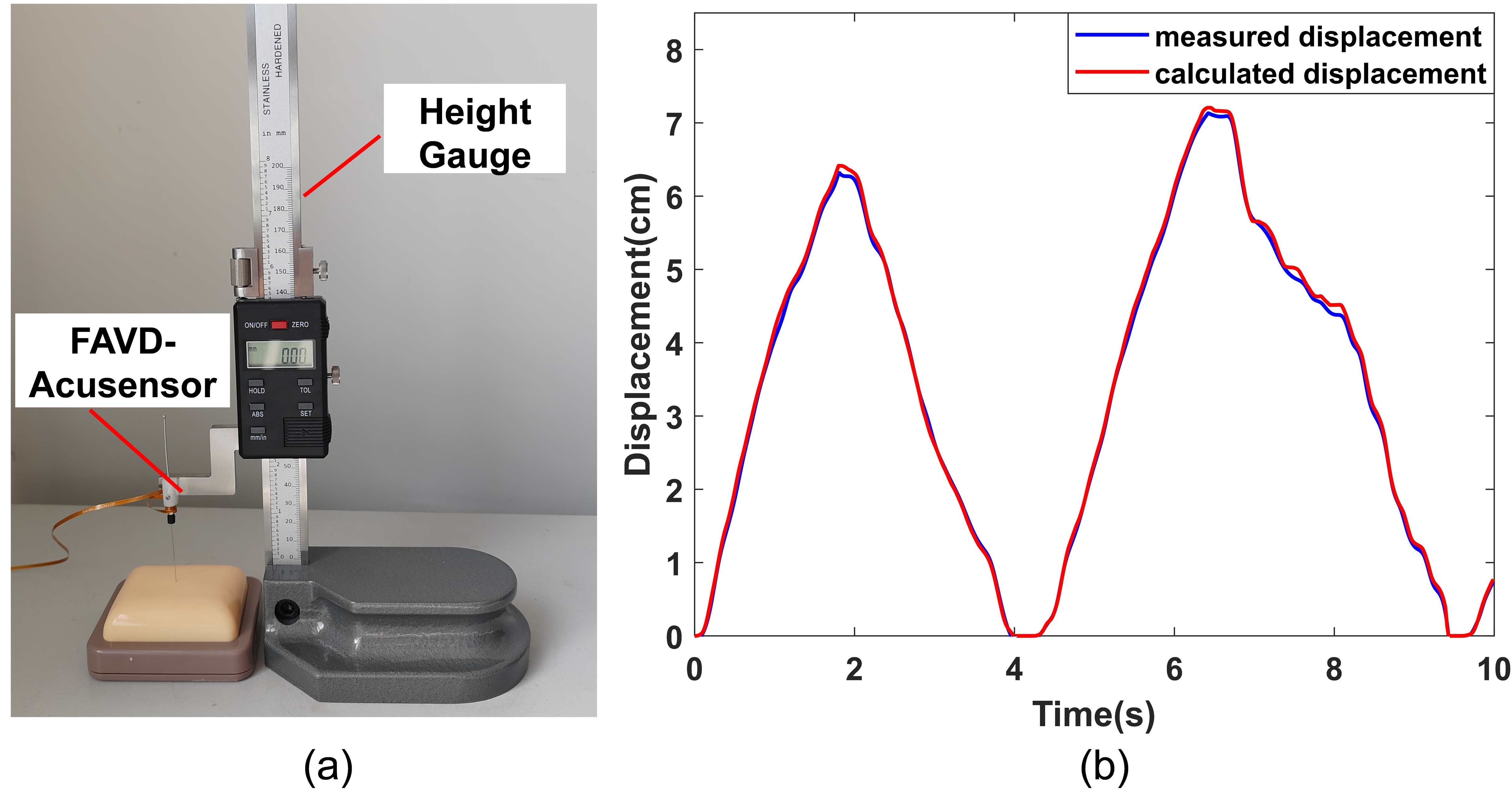}
\caption{Testing of displacement. (a) Experimental setup for the displacement measurement. (b) Comparison between calculated displacement and measured displacement.}
\label{fig_10}
\end{figure}

\begin{figure*}[t]
\centering
\includegraphics[width=0.85\linewidth]{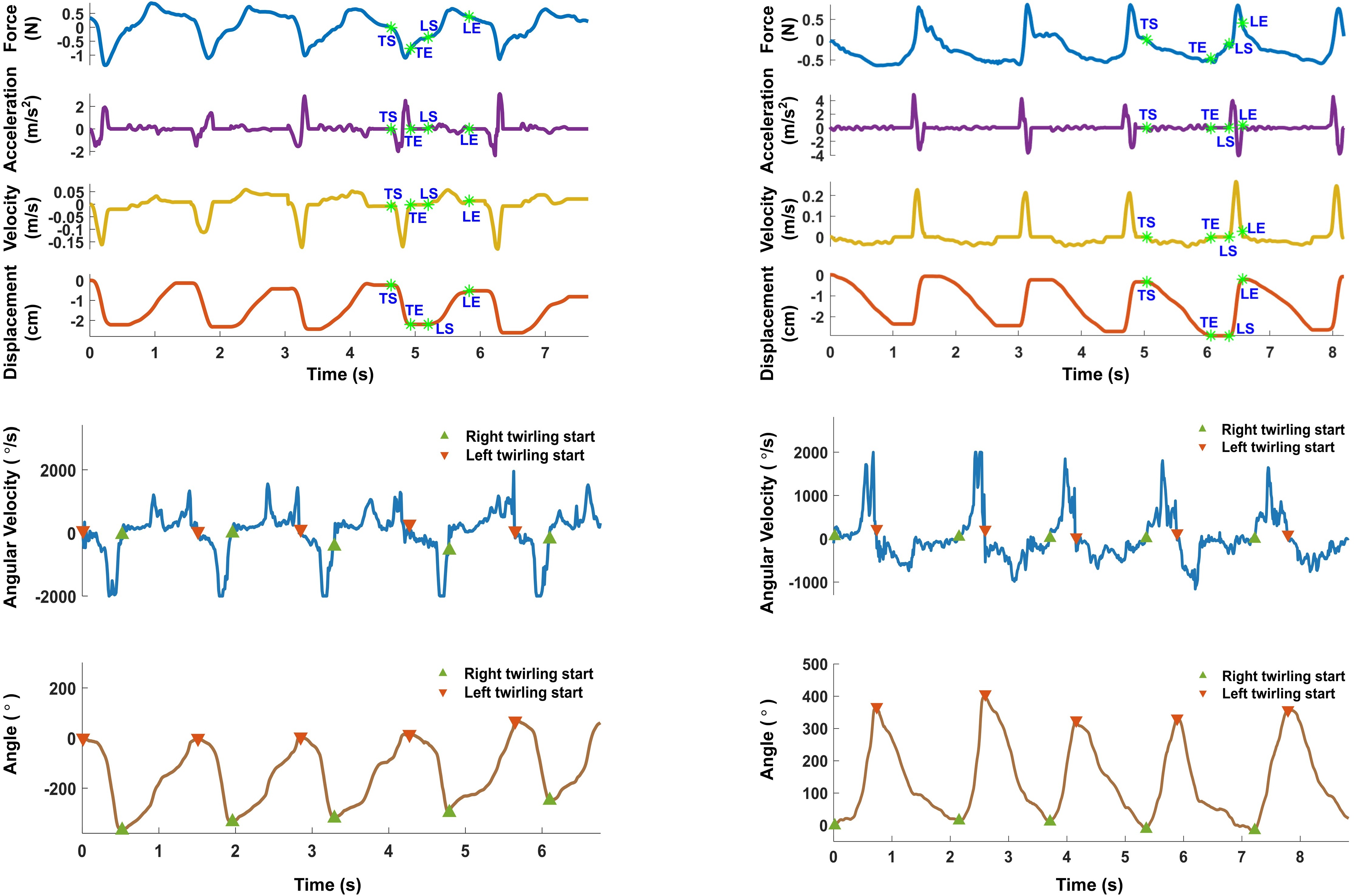}
\caption{Waveforms of key parameters of four basic acupuncture manipulations. (a) The waveform of LTRF manipulation. (b) The waveform of LTRD manipulation. (c) The waveform of TRRF manipulation. (d) The waveform of TRRD manipulation.}
\label{fig_11}
\end{figure*}

Firstly, we investigate the mechanical features of LTRF and LTRD in lifting-thrusting operations.  The coordinates of the four characteristic points (TS, RE, LS, LE) are annotated in the acceleration, force, velocity, and displacement waveforms of the lifting-thrusting operation.  Fig. 11(a) illustrates the waveforms of five complete cycles of the LTRF. 
As shown in Fig. 11(a), within each cycle, the LTRF is characterized by relatively high velocity, force, and acceleration during the thrusting stage, while the lifting stage exhibits relatively low velocity, force, and acceleration, with more gradual variations. As shown in Fig. 11(b), the waveforms for the LTRD reveal the following characteristics during each cycle: the thrusting stage is characterized by a relatively slow velocity and low, gradual changes in force and acceleration, whereas the lifting stage exhibits a relatively fast velocity with higher, more rapid changes in force and acceleration. As shown in Fig. 11(a) and (b), the start point of the thrusting stage is neither the numerical zero point mentioned in \cite{ref32} nor the waveform rising start point mentioned in \cite{ref33}. This also indicates that defining lifting-thrusting operation cycles solely based on applied force has certain limitations, while defining cycles based on acceleration provides a more accurate approach. 

The mechanical features of TRRF and TRRD manipulations during the twirling-rotating operation are analyzed. Fig. 11(c) presents the angular velocity and angle waveform diagrams of five cycles for the TRRF, which includes two stages: (1) thumb progressing forward (needle twirling to left) with greater torque; (2) thumb regressing backward (needle twirling to right) with lesser torque. During the first stage, the angular velocity is negative, leading to a decrease in the twirling-rotating angle; the greater torque applied is reflected in a shorter operation time and a higher root mean square (RMS) value of the angular velocity. During the second stage, the angular velocity becomes positive, the lesser torque applied is indicated by a longer operation time and a lower RMS value of the angular velocity. Fig. 11(d) illustrates the waveform diagram of the TRRD. The operational characteristics involve exerting heavy torque with the thumb regressing backward and light torque with the thumb progressing forward, contrary to the TRRF manipulation. The TRRD manipulation commences with the thumb regressing backward followed by the thumb progressing forward. When the needle twirling to right, the angle rapidly increases; when it twirling to left, the angle slowly decreases.

\begin{figure*}[t]
\centering
\includegraphics[width=\linewidth]{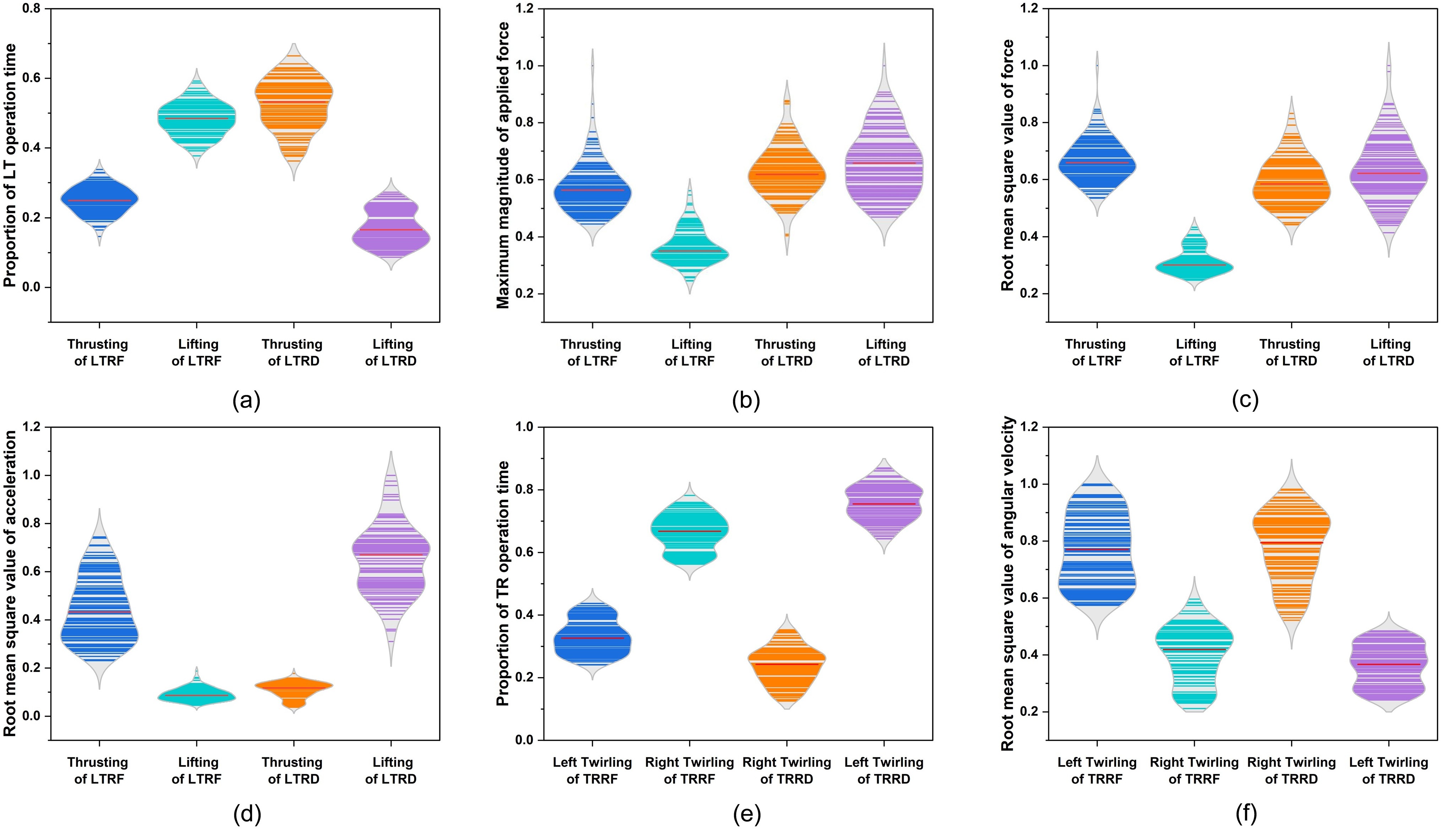}
\caption{Violin plots show features in different manipulations and stages of lifting-thrusting operation and twirling-rotating operation. (a) Proportion of LT operation time. (b) Maximum magnitude of applied force. (c) Root mean square value of force. (d) Root mean square value of acceleration. (e) Proportion of TR operation time. (f) Root mean square value of angular velocity.\\
LT: lifting-thrusting, TR: twirling-rotating. 
}
\label{fig_12}
\end{figure*}

\begin{table}[t]
\centering
\caption{Features of Acupuncture Manipulations}
\begin{tabular}{@{}cc@{}}
\toprule
Feature Number & Feature Description \\
\midrule
F1 & Proportion of lifting--thrusting operation time \\
F2 & Maximum magnitude of applied force \\
F3 & Root mean square value of force \\
F4 & Root mean square value of acceleration \\
F5 & Proportion of twirling--rotating operation time \\
F6 & Root mean square value of angular velocity \\
\bottomrule
\end{tabular}
\end{table}

To further quantitatively describe different acupuncture manipulations, six features (F1-F6) representing the operational patterns of the lifting-thrusting and twirling-rotating operations were defined as shown in Table IV. Among these, F1 represents the proportion of operation time for downward thrusting and upward lifting within their respective single cycles; F2 reflects the maximum magnitude of applied force during the acupuncture process and can indicate its intensity. F3 (Root mean square value of force) and F4 (Root mean square value of acceleration) can evaluate the overall energy magnitude of force and acceleration, respectively. F5 represents the proportion of operation time for left-ward twirling or right-ward twirling within a single cycle, while F6 (Root mean square value of angular velocity) can evaluate the overall energy magnitude of angular velocity. An analysis of variance using the Welch test was conducted on the F1-F4 features of 200  lifting-thrusting cycles and the F5-F6 features of 200 twirling-rotating cycles to compare inter-group differences among different manipulations and stages. $P<0.05$ was used as the significance threshold, and results are expressed as Mean±standard deviation (SD).

\begin{table}[t]
\centering
\caption{Comparison of FEATURES of different manipulations and stages of lifting–thrusting operation}
\begin{tabular}{@{}ccccc@{}}
\toprule
\makecell{Feature \\Number} & \makecell{Thrusting \\of LTRF} & \makecell{Lifting \\of LTRF} & \makecell{Thrusting \\of LTRD} & \makecell{Lifting \\of LTRD} \\
\midrule
F1 & 0.25 $\pm$ 0.08 & 0.48 $\pm$ 0.05 & 0.52 $\pm$ 0.07 & 0.17 $\pm$ 0.05 \\
F2 & 0.67 $\pm$ 0.08 & 0.32 $\pm$ 0.05 & 0.60 $\pm$ 0.08 & 0.64 $\pm$ 0.12 \\
F3 & 0.58 $\pm$ 0.09 & 0.37 $\pm$ 0.06 & 0.63 $\pm$ 0.09 & 0.63 $\pm$ 0.12 \\
F4 & 0.45 $\pm$ 0.14 & 0.091 $\pm$ 0.03 & 0.11 $\pm$ 0.03 & 0.65 $\pm$ 0.15 \\
\bottomrule
\end{tabular}
\end{table}

Table V presents the Mean±SD of the four features (F1-F4) for LTRF and LTRD. Fig. 12 illustrates the violin plots of feature characteristics for different lifting-thrusting operations and stages using F1-F4. The following conclusions can be drawn:

(1) The F1, F2, F3, and F4 features demonstrate significant differences between the downward thrusting and upward lifting stages of the LTRF ($P<0.05$). This suggests that these features objectively characterize the characteristics of LTRF manipulation, as greater force is applied during downward thrusting and less during upward lifting.

(2) The F1, F2, and F4 features demonstrate significant differences between the downward thrusting and upward lifting stages of the LTRD ($P<0.05$). This suggests that these features objectively characterize the LTRD manipulation, as lighter force is applied during downward thrusting and greater force during upward lifting. In contrast, the F3 feature fails to show significant differences ($P > 0.05$), indicating that acceleration-based features may be more advantageous than force-based features for objectively quantifying acupuncture manipulation.

(3) The F1, F2, F3, and F4 features demonstrate significant differences ($P < 0.05$) during both the thrusting and lifting stages between LTRF and LTRD manipulations. This indicates that these features are appropriate for classifying acupuncture manipulations in research studies.

(4) Despite being characterized by rapid changes, significant differences ($P<0.05$) were observed in the F1, F3, and F4 features between the downward thrusting stage of LTRF and the upward lifting stage of LTRD. Notably, no significant difference ($P > 0.05$) was found in the F2 feature. Furthermore, the F4 feature during the upward lifting stages of LTRD was significantly higher than that during the downward thrusting stage of LTRF, despite no significant difference in peak force. This discrepancy may be attributed to the greater frictional force and needle tip force experienced during the downward thrusting stage compared to the upward lifting stage. These findings further highlight that acceleration signals are more effective than force signals in distinguishing different acupuncture manipulations.

\begin{table}[t]
\centering
\caption{Comparison of FEATURES of different manipulations and stages of twirling-rotating operation}
\begin{tabular}{@{}ccccc@{}}
\toprule
\makecell{Feature \\Number} & \makecell{Left Twirling \\
of TRRF} & \makecell{Right twirling \\
of TRRF} & \makecell{Right twirling \\
of TRRD} & \makecell{Left Twirling \\
of TRRD} \\
\midrule
F5 & 0.33 $\pm$ 0.05 & 0.66 $\pm$ 0.06 & 0.24 $\pm$ 0.06 & 0.76 $\pm$ 0.06 \\
F6 & 0.77 $\pm$ 0.13 & 0.40 $\pm$ 0.10 & 0.78 $\pm$ 0.12 & 0.36 $\pm$ 0.07 \\
\bottomrule
\end{tabular}
\end{table}

Table VI presents the Mean±SD of two features (F5 and F6) for the TRRF and TRRD manipulations. Fig. 12(e) and (f) display the violin plots for the TRRF and TRRD manipulations and their respective stages, supporting the following conclusions:

(1) The F5 and F6 features demonstrate significant differences ($P<0.05$) during both the left-ward and right-ward twirling stages in TRRF and TRRD. These findings objectively characterize the operational characteristics of the manipulations. In TRRF, the thumb forward stage is characterized by rapid rotation while the thumb backward stage is characterized by slower rotational motion. Conversely, in TRRD, the thumb backward stage exhibits rapid rotation, while the thumb forward stage exhibits slower rotation.

(2) The F5 and F6 features demonstrate significant differences ($P<0.05$) during the left-ward and right-ward twirling stages in TRRF and TRRD. This indicates that the extracted parameter features are appropriate for classifying different twirling-rotating operations.

\section{DISCUSSION AND CONCLUSION }
\subsection{Discussion}
 This study makes two significant contributions. First, it establishes the kinematic and dynamic model of acupuncture, which collectively form a comprehensive mechanical framework for the entire acupuncture procedures. This framework offers a systematic explanation of the interrelationships among the kinematic and dynamic parameters involved in acupuncture. Second, the study proposes an innovative sensing acupuncture needle named FAVD-Acusensor integrating an IMU and a force sensor. Through a multimodal signal processing chain, key parameters in the kinematic and dynamic model, including lifting-thrusting force, acceleration, velocity, displacement, as well as twirling-rotating angular velocity and angle, are resolved. The methodology emphasizes the calculation of lifting-thrusting velocity and displacement via a fusion method combining visual and IMU data, addressing the challenge of detecting these parameters during practical acupuncture procedures. The quantification of key parameters elucidates the differences in basic acupuncture manipulations, enabling the objective characterization of acupuncture operations. This facilitates standardized instruction, providing clear references for novice apprentices and laying the foundation for the preservation and transmission of acupuncture therapy. 
 
Compared with methods relying solely on force signal analysis, the proposed approach in this study offers several advantages. (1) Comprehensive Information. Integrating IMU and force sensing provides additional kinematic parameters, such as lifting-thrusting velocity and displacement, which are otherwise difficult to measure during acupuncture procedures. Lifting-thrusting velocity and displacement are critical kinematic parameters in acupuncture. They play a pivotal role in regulating nerve function and are closely linked to therapeutic outcomes. These parameters serve as valuable feedback on the operational characteristics of students, indicating whether the thrusting velocity is within an appropriate range or if the displacement meets the required standard. This enables students to effectively rectify errors and improve their skills. (2) Significant differences exist between the manipulations.  In the comparative evaluation of different acupuncture manipulations and stages, acceleration signals more effectively capture the mechanical characteristics and operational nuances compared to force signals. (3) Accurate cycle segmentation is achieved. The acceleration and visual-based cycle segmentation method incorporates the needle tip force and the frictional force. Compared to the traditional approach of obtaining it from the force signal, this novel method exhibits superior dynamic behavior and enhanced feature recognition precision. (4) Evaluation of the forces exerted by the tissues. Through a comprehensive analysis of the characteristics of acceleration and applied force, the biomechanical interaction between biological tissues and the needle enables a quantitative evaluation. This approach allows for the assessment of the forces involved, as well as the relationship between the needle tip force and the frictional force exerted by the tissues on the needle during different lifting-thrusting stages.

Within the framework of the FAVD-Acusensor and acupuncture quantitative system, we can systematically identify numerous characteristics of acupuncture mechanics. Through research on the operation characteristics and mechanical properties of the four basic acupuncture manipulations, we achieve a significant advancement in the accurate quantification of acupuncture parameters. Based on these characteristics, we are able to distinguish the differences between acupuncture manipulations and provide an objective description of the patterns associated with acupuncture operation. The synergistic observation of diverse signals during acupuncture enables a deeper analysis of the biomechanical mechanisms underlying tissue responses. The analysis of lifting-thrusting force and acceleration during the Deqi sensation (the sensation of the ``acupuncture effect") provides insight into the mechanical characteristics associated with effective acupuncture procedures.   

\subsection{Limitations and Future Work}
(1) In this study, we employed a monitoring method based on angular velocity to effectively assess the rotational characteristics of the needle during twirling-rotating operation, thereby avoiding the volume constraints of torque sensors and significantly reducing the size of the sensing needle. However, this method fundamentally reflects the dynamic state of the needle through indirect means. In scenarios such as acupuncture robots, where precise torque information is required, direct torque measurement remains indispensable. Therefore, future research will further analyze and validate torque estimation methods based on angular velocity to achieve a more accurate assessment of the needle's dynamic state while maintaining the miniaturization of the sensing needle. 

(2) This study conducted a systematic experimental validation of four basic acupuncture manipulations. Given their relatively low clinical utilization frequency, manipulations such as flicking and trembling were excluded from the present experimental scope. In future prospective studies, we aim to achieve precise calculation and quantification of three-dimensional angular velocity and angles for special manipulations such as flicking and trembling by employing coordinate transformation between the body-fixed and inertial coordinate systems. This approach will further enhance the capabilities of the FAVD-Acusensor in quantifying various acupuncture manipulations.

Through the application of the FAVD-Acusensor proposed in this study, it is possible to rigorously analyze the mechanical characteristics of the four basic acupuncture manipulations. This innovation provides a tool for achieving standardized measurement of acupuncture stimulus intensity and for advancing clinical research into the relationship between acupuncture dosage and therapeutic efficacy.
\subsection{Conclusion}
To investigate the quantification of acupuncture manipulations, we have developed the FAVD-Acusensor capable of detecting kinematic and dynamic parameters, along with an acupuncture quantitative system that provides comprehensive manipulation information. The FAVD-Acusensor measures the lifting-thrusting parameters, including force, acceleration, velocity, and displacement, as well as the twirling-rotating parameters, including angular velocity and angle, with a weight of only 1.3 grams. The acupuncture quantitative system offers advantages over traditional systems through its compact design, lightweight construction, and high precision (a nonlinearity error of ±0.45\% in force measurement and a displacement RMSE of 1.2 mm). Experimental results demonstrate significant differences among the four basic acupuncture manipulations, indicating the system's ability to accurately capture force and motion information. This study provides valuable tools and methods to advance standardized acupuncture education and optimize trajectories for acupuncture robots.


\begin{thebibliography}{1}
\bibliographystyle{IEEEtran}

\bibitem{ref1}
``WHO benchmarks for the practice of acupuncture: CC BY-NC-SA 3.0 IGO," World Health Organization, Geneva, 2020. 

\bibitem{ref2}
L. Lu et al., ``Evidence on acupuncture therapies is underused in clinical practice and health policy," \textit{The British Medical Journal}, vol. 376, p. e067475, Feb 25, 2022.

\bibitem{ref3}
Y. Wang, X. Shi, T. Efferth, and D. Shang, ``Artificial intelligence-directed acupuncture: a review," \textit{Chinese Medicine}, vol. 17, no. 1, 2022. 
\bibitem{ref4}
R. Lyu, M. Gao, H. Yang, Z. Wen, and W. Tang, ``Stimulation Parameters of Manual Acupuncture and Their Measurement," \textit{Evid Based Complement Alternat Med}, vol. 2019, p. 1725936, 2019.

\bibitem{ref5}
W. T. H. Ji, T. Wu, W. Shu, X. Wu, X. Ya, C. Ji, X. Wang and Y.Han, ``The Mechanics Basis of Acupuncture Therapy," \textit{Applied Mathematics and Mechanics}, vol. 45(6), pp. 803-822, 2024. 

\bibitem{ref6}
Q. Liang et al., ``FBG-FT-IMU: Multimodal Measurement System for Acupuncture Manipulation," \textit{IEEE Transactions on Instrumentation and Measurement}, vol. 74, pp. 1-10, 2025.

\bibitem{ref7}
G. Du, Y. Li, K. Su, C. Li, and P. X. Liu, ``A Mobile Natural Human–Robot Interaction Method for Virtual Chinese Acupuncture,"  \textit{IEEE Transactions on Instrumentation and Measurement}, vol. 72, pp. 1-10, 2023.

\bibitem{ref8}
M. Z. Sun, X. Wang, Y. C. Li, W. Yao, and W. Gu, ``Mechanical effects of needle texture on acupoint tissue,"  \textit{J Integr Med}, vol. 21, no. 3, pp. 254-267, May 2023.

\bibitem{ref9}
Q. Liang, S. Ouyang, J. Long, L. Zhou, and D. Zhang, ``A Novel Fiber Bragg Grating Three-Dimensional Force Sensor for Medical Robotics," \textit{IEEE/ASME Transactions on Mechatronics}, pp. 1-11, 2024.

\bibitem{ref10}
J. Xu and A. Song, ``A Miniature Multiaxis Force/Torque Sensor for Acupuncture," \textit{IEEE Sensors Journal}, vol. 23, no. 7, pp. 6660-6671, 2023.

\bibitem{ref11}
H.-y. Yang, Y.-y. Liu, M. Gao, H. Yin-e, and Z. Hu, 
J. Xu and A. Song, ``Research on Real-time Collection of Acupuncture Manipulation Parameter and Teaching Demonstration System," \textit{presented at the 2009 WRI World Congress on Software Engineering}, 2009.

\bibitem{ref12}
Y. J. Han, S. Y. Yi, Y. J. Lee, K. H. Kim, E. J. Kim, and S. D. Lee, ``Quantification of the parameters of twisting-rotating acupuncture manipulation using a needle force measurement system," \textit{Integr Med Res}, vol. 4, no. 2, pp. 57-65, Jun 2015.

\bibitem{ref13}
R. T. Davis, D. L. Churchill, G. J. Badger, J. Dunn, and H. M. Langevin, ``A new method for quantifying the needling component of acupuncture treatments," \textit{Acupunct Med}, vol. 30, no. 2, pp. 113-9, Jun 2012.

\bibitem{ref14}
H. Yu, Z. Zhu, C. Wang, J. Wang, and C. Liu,  ``Kinematic-driven human-robot interaction system with deep learning for flexible acupuncture needling manipulations," \textit{Biomedical Signal Processing and Control}, vol. 92, 2024.

\bibitem{ref15}
J. Su, Y. Zhu, and M. Zhu, ``Hand-Eye-Force Coordination of Acupuncture Robot," \textit{IEEE Access}, vol. 7, pp. 82154-82161, 2019.

\bibitem{ref16}
M. Muthukumar, M. S. Bobji, and K. R. Y. Simha, ``Needle insertion-induced quasiperiodic cone cracks in hydrogel," \textit{Soft Matter}, vol. 17, no. 10, pp. 2823-2831, Mar 18 2021

\bibitem{ref17}
S. Ocal, M. U. Ozcan, I. Basdogan, and C. Basdogan, ``Effect of preservation period on the viscoelastic material properties of soft tissues with implications for liver transplantation," \textit{J Biomech Eng}, vol. 132, no. 10, p. 101007, Oct 2010.

\bibitem{ref18}
D. P. Widianto et al., ``Microneedle Insertion into Visco-Hyperelastic Model for Skin for Healthcare Application," \textit{presented at the 2021 IEEE 71st Electronic Components and Technology Conference (ECTC)}, 2021.

\bibitem{ref19}
W. Yao, Z. Shen, Y. Yu, and G. Ding, ``Mechanical effects of acupuncture," \textit{Mathematical Methods in the Applied Sciences}, vol. 43, no. 4, pp. 1555-1564, 2019.

\bibitem{ref20}
Y. Yu, W. Yao, and G. Ding, ``A Mathematical Model to Study the Mechanical Information Induced by Lifting-Thrusting Needle," \textit{Evid Based Complement Alternat Med}, vol. 2019, p. 5475426, 2019.

\bibitem{ref21}
W. Yao, Y. Yu, and G. Ding, ``A hybrid method to study the mechanical information induced by needle rotating," \textit{Mathematical Methods in the Applied Sciences}, vol. 41, no. 15, pp. 5939-5950, 2018.

\bibitem{ref22}
L. Yunshan, X. Chengli, Z. Peiming, Q. Haocheng, L. Xudong, and L. Liming, ``Integrative research on the mechanisms of acupuncture mechanics and interdisciplinary innovation," \textit{Biomed Eng Online}, vol. 24, no. 1, p. 30, Mar 7 2025.

\bibitem{ref23}
C. Shi, Z. Tang, and S. Wang, ``Design and Experimental Validation of a Fiber Bragg Grating-Enabled Force Sensor With an Ortho-Planar Spring-Based Flexure for Surgical Needle Insertion," \textit{IEEE Transactions on Medical Robotics and Bionics}, vol. 3, no. 2, pp. 362-371, 2021.

\bibitem{ref24}
C. Su, C. Wang, S. Gou, J. Chen, W. Tang, and C. Liu, ``An action recognition method for manual acupuncture techniques using a tactile array finger cot," \textit{Comput Biol Med}, vol. 148, p. 105827, Sep 2022.

\bibitem{ref25}
B. Chen, K. Lin, L. Xu, J. Cao, and S. Gao, ``A Piezoelectric Force Sensing and Gesture Monitoring-Based Technique for Acupuncture Quantification," \textit{IEEE Sensors Journal}, vol. 21, no. 23, pp. 26337-26344, 2021.

\bibitem{ref26}
J. Li, L. Grierson, M. X. Wu, R. Breuer, B. Zhou, and H. Carnahan, ``Research on action features of acupuncturist experts' acupuncture needle twirling skills," \textit{Acupuncture research}, vol. 38, no. 05, pp. 415-419, 2013.

\bibitem{ref27}
H. Fourati, ``Heterogeneous Data Fusion Algorithm for Pedestrian Navigation via Foot-Mounted Inertial Measurement Unit and Complementary Filter," \textit{IEEE Transactions on Instrumentation and Measurement}, vol. 64, no. 1, pp. 221-229, 2015.

\bibitem{ref28}
J. Chang, J. Cieslak, J. Davila, J. Zhou, A. Zolghadri, and Z. Guo, ``A Two-Step Approach for an Enhanced Quadrotor Attitude Estimation via IMU Data," \textit{IEEE Transactions on Control Systems Technology}, vol. 26, no. 3, pp. 1140-1148, 2018.

\bibitem{ref29}
Q. Xu, P. Hu, W. Li, and J. Qiao, ``System Modeling and Adaptive Strong-Tracking Filter Design for Underwater Polarization Navigation Against Variable Measurement Bias," \textit{IEEE Transactions on Industrial Electronics}, pp. 1-10, 2025.

\bibitem{ref30}
H. Liu, X. Sun, and S. Bai, ``Modified Unscented Kalman Filter Considering Maximum Point of Probability Density Function," \textit{IEEE Transactions on Aerospace and Electronic Systems}, vol. 61, no. 2, pp. 4926-4944, 2025.

\bibitem{ref31}
‌Kalman, R. E, ``A New Approach to Linear Filtering and Prediction Problems," \textit{Journal of Basic Engineering}, vol. 82, no. 1, pp. 35-45, 1960.

\bibitem{ref32}
J. Xu, A. Song, and Y. Chen, ``Acupuncture Manipulation Measurement and Identification Based on Miniature Multiaxis Force/Torque Sensor," \textit{IEEE Sensors Journal}, vol. 24, no. 20, pp. 33806-33815, 2024.

\bibitem{ref33}
G. Ding, X. Shen, Y. Tao, H. Liu, W. Yao, and X. Li, ``The contrast research between acupuncture manipulations and the parameters of force acting on needle," \textit{Chinese Journal of Biomedical Engineering}, no. 04, pp. 334-341, 2004.

\bibitem{ref34}
D. W. Allan, "Statistics of atomic frequency standards, ``\textit{in Proceedings of the IEEE}, vol. 54, no. 2, pp. 221-230, 1966.
\end{thebibliography}
\end{document}